\documentclass[fleqn,10pt]{wlscirep}
\usepackage[utf8]{inputenc}
\usepackage[T1]{fontenc}
\usepackage{bm}
\usepackage{xcolor}
\usepackage{setspace}
\usepackage{ulem}
\usepackage{float}

\title{Atomistic Fracture in bcc Iron Revealed by Active Learning of Gaussian Approximation Potential}

\author[1,*]{Lei Zhang}
\author[2]{Gábor Csányi}
\author[3]{Erik van der Giessen}
\author[1,*]{Francesco Maresca}
\affil[1]{Engineering and Technology Institute, Faculty of Science and Engineering,
	University of Groningen, Nijenborgh 4, 9747 AG Groningen, The Netherlands}
\affil[2]{Engineering Laboratory, University of Cambridge, Cambridge CB2 1PZ, United Kingdom}
\affil[3]{Zernike Institute for Advanced Materials, University of Groningen, Nijenborgh 4,
	9747 AG Groningen, The Netherlands}

\affil[*]{correspondence to: lei.zhang@rug.nl, f.maresca@rug.nl}

\begin{abstract}
The prediction of atomistic fracture mechanisms in body-centred cubic (bcc) iron is essential for understanding its semi-brittle nature.
Existing atomistic simulations of the crack-tip deformation mechanisms under mode-I loading based on classical interatomic potentials yield contradicting predictions. 
To enable fracture prediction with quantum accuracy, we develop a Gaussian approximation potential (GAP) using an active learning strategy by extending a density functional theory (DFT) database of ferromagnetic bcc iron. 
We apply the active learning algorithm and obtain a Fe GAP model with a maximum predicted error of 8 meV/atom over a broad range of stress intensity factors (SIFs) and for four crack systems. 
The learning efficiency of the approach is analysed, and the predicted critical SIFs are compared with Griffith and Rice theories.
The simulations reveal that cleavage along the original crack plane is the crack tip mechanism for $\{100\}$ and $\{110\}$ crack planes at T=0K, thus settling a long-standing dispute.
Our work also highlights the need for a multiscale approach to predicting fracture and intrinsic ductility, whereby finite temperature, finite loading rate effects and pre-existing defects (e.g. nanovoids, dislocations) should be taken explicitly into account.
\end{abstract}
\begin{document}
	
\setstretch{1.5}
\flushbottom
\maketitle

\thispagestyle{empty}

\section*{Introduction}

\begin{figure}[H]
	\centering
	\includegraphics[trim=0 0 0 0, clip, scale=0.4]{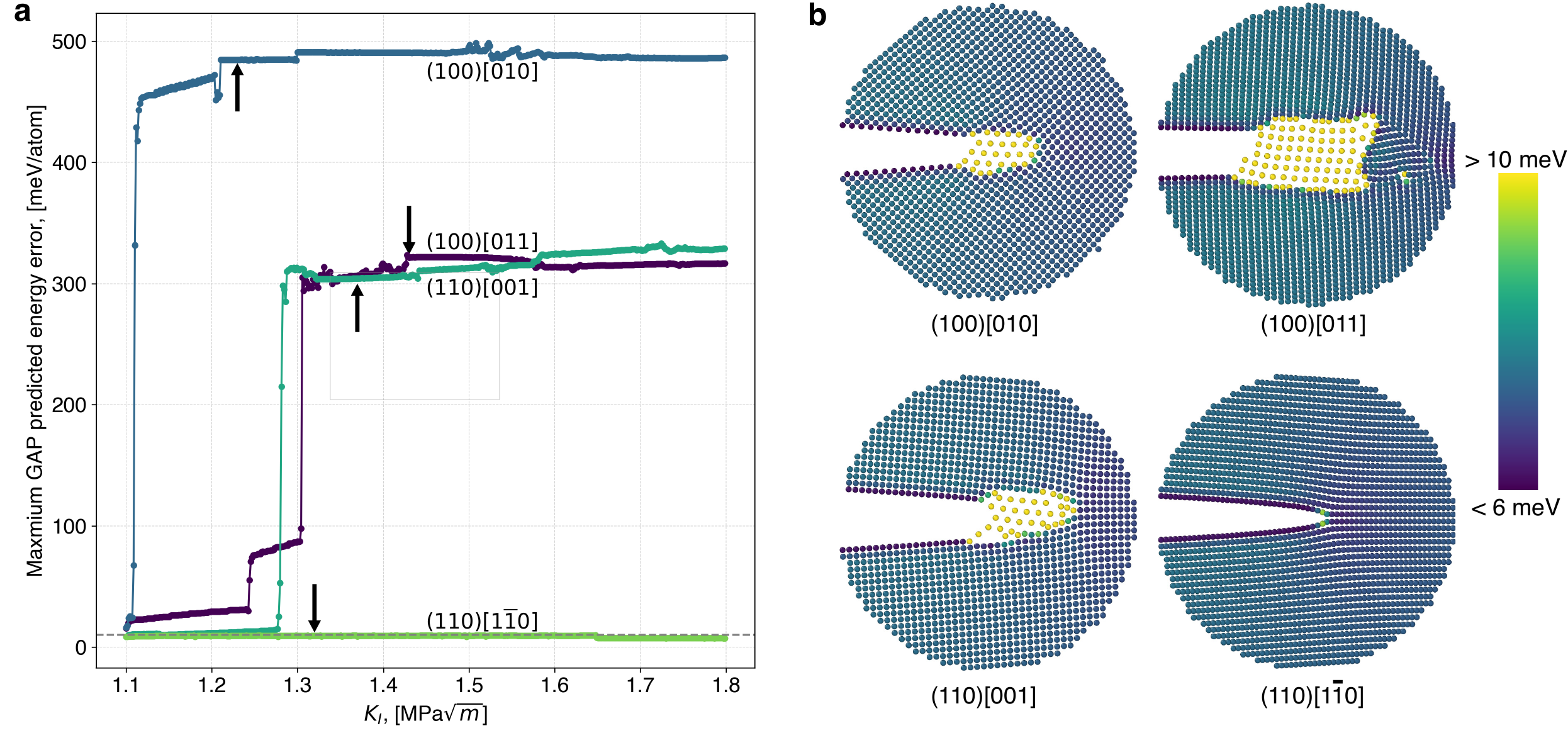}
	\caption{Fracture predictions of Fe-GAP18 trained on Dragoni \textit{et al.} original database\cite{Dragoni2018_PhysRevMaterials.2.013808}. \textbf{(a)} Maximum per-atom GAP predicted energy error during fracture simulations of four crack systems. The dashed line indicates the 10 meV/atom, and the arrows indicate the configuration shown in \textbf{(b)}. \textbf{(b)} Simulation snapshots coloured by GAP predicted per-atom energy error. Note that this is a new version of Fe-GAP18 trained on the database by Dragoni \textit{et al.} using an optimised descriptor\cite{turboSOAP_PhysRevB.100.024112} (see section \textit{Methods: GAP training}).
	}
	\label{fig:old_gap}
	\vspace*{-5mm}
\end{figure}

 Brittle fracture is a key failure mechanism of body-centred cubic (bcc) transition metals, which limits their application and can jeopardise the safety of infrastructures. 
Brittle fracture usually takes place in the form of cleavage, which can be accompanied by dislocation plasticity. 
The competition between thermally activated dislocation mobility and (atomic-scale) crack tip deformation mechanisms controls the fracture process of bcc iron \cite{Gumbsch_2002,RICE1992239,Andric_2018,MAK2021104389}.
Experiments on single-crystal bcc iron reveal that cleavage takes place on $\{100\}$ planes for pre-existing $\{100\}$, and $\{110\}$ crack planes with $\langle 100 \rangle$ and $\langle 110 \rangle$  crack fronts\cite{Hribernik_thesis}. 
However, atomic-scale experimental data is unavailable for crack tips in bcc iron, leaving the atomistic crack-tip mechanisms unclear.\\
\indent Atomistic modelling based on molecular dynamics (MD) approach has been widely used to predict the crack-tip mechanisms\cite{deCelis_1983,Guo_2007_acta,Guo_2006,Guo_2007_msea,cao_wang_2006,Wang_2021,Meiser_2020,Cheung1994AMS,Gordon2007CrackTipDM,Moller_2014}(see \textit{Supplementary Material S1} for a summary of existing studies). 
Mode-I atomistic crack tip simulations at T=0K are typically used to assess the intrinsic ductility of metals, which is controlled by the competition between crack propagation (cleavage) and dislocation emission from the crack tip\cite{kermode2008low,Andric_2018,MAK2021104389}.
Yet, as highlighted in extensive reviews\cite{Gordon2007CrackTipDM,Moller_2014}, classical interatomic potentials (IAPs) for bcc iron predict contradicting the crack-tip mechanisms at T=0K (i.e. cleavage, crack propagation planes, dislocation emission, and phase transition) for the same crack system.
Remarkably, none of them agree with the low-temperature experimental fracture mechanisms as a function of the pre-existing crack system\cite{Hribernik_thesis}.
\begin{figure}[H]
	\centering
	\includegraphics[ trim=0 0 0 85, clip, scale=0.42]{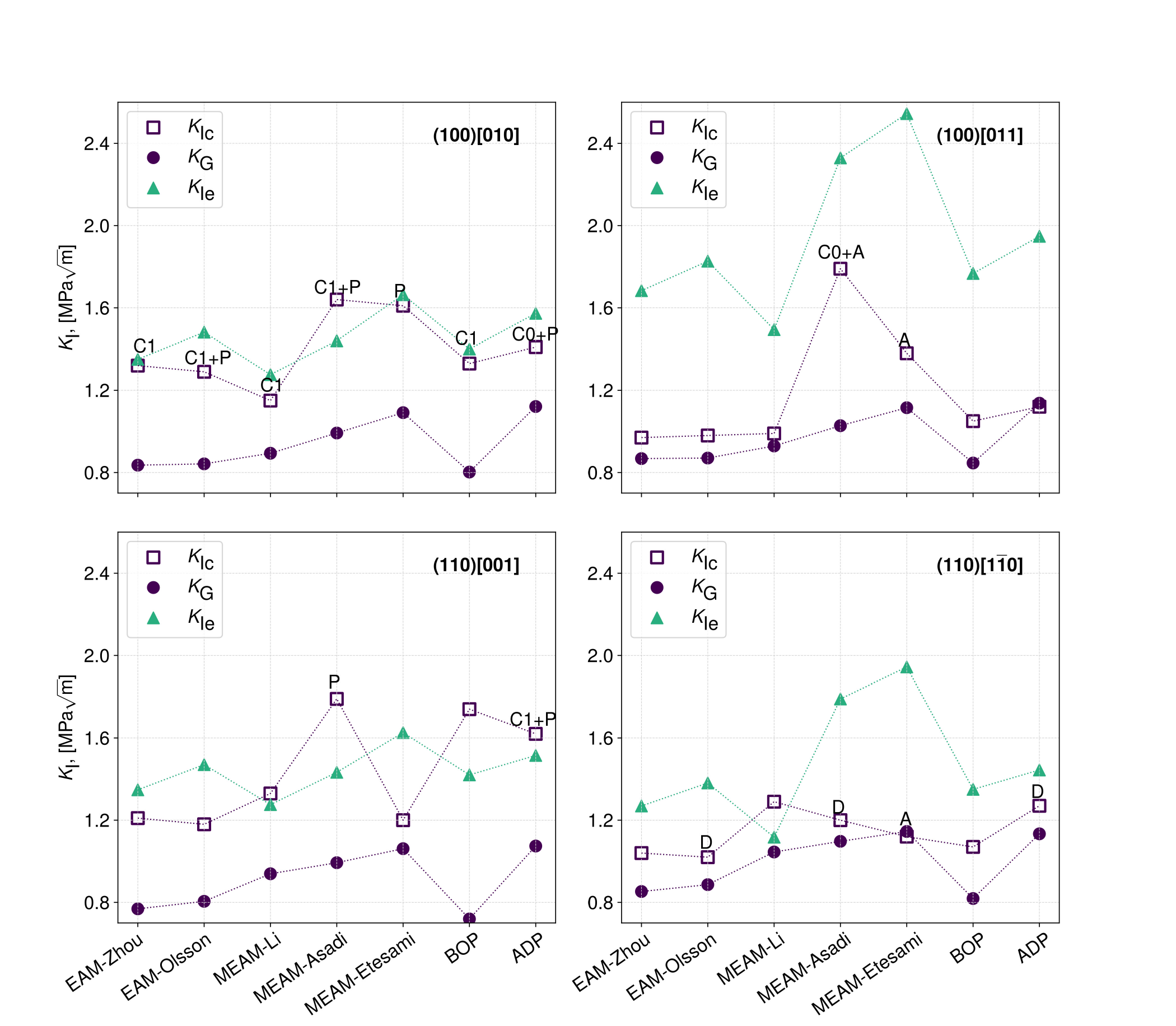}
	\caption{Summary of fracture simulations for crack system $(100)[010]$, $(100)[011]$, $(110)[001]$, and $(110)[1\overline{1}0]$. 
		$K_{\rm Ic}$ is the critical stress intensity factor predicted by MS simulations. 
		The letters above the symbols indicate the fracture mechanism observed in MS simulations with the different IAPs. 
		\textbf{C0} and \textbf{C1} indicate cleavage on \{100\} and \{110\} planes that are different from the original crack plane. 
		\textbf{P} and \textbf{D} indicate phase transformation and dislocation emission from the crack tip, respectively. 
		\textbf{A} indicates amorphous structure forming at the crack tip. 
		The crack propagates on the original crack plane if no symbol is specified.
		$K_{\rm G}$ and $K_{\rm Ie}$ are predictions according to Griffith\cite{Griffith} and Rice 	theories\cite{RICE1992239} for cleavage and dislocation emission, respectively.}
	\label{fig:iap_after2015}
	\vspace*{-5mm}
\end{figure}
\indent Fig. \ref{fig:iap_after2015} shows the critical stress intensity factor ($K_{\rm Ic}$) under mode-I loading as computed by performing molecular statics (MS) simulations using seven classical IAPs (see \textit{MS/MD simulation setup} in \textit{Methods}) that were not investigated in Ref.\cite{Gordon2007CrackTipDM} nor Ref.\cite{Moller_2014}, including two embedded atom method (EAM) potentials\cite{zhou_2004_PhysRevB.69.144113,olsson_2009}, three Modified EAM (MEAM) potentials\cite{meam_liyange_PhysRevB.8pg
	9.094102,Asadi_2015,Asadi_2018}, one bond order potential (BOP)\cite{Byggmastar_2020}, and one angular dependent potential (ADP)\cite{adp_PhysRevMaterials.5.063607}.
Results are compared with the critical stress intensity factors according to Griffith theory \cite{Griffith} 
	\begin{equation}
		K_{\rm G}=\sqrt{\frac{2\gamma_{\rm s}}{B}},
	\end{equation}
where $\gamma_{\rm s}$ is the surface energy and $B$ is a constant determined by the elastic constants (see \textit{Supplementary Material S2.2}). 
Dislocation emission is predicted according to Rice theory\cite{RICE1992239} with an anisotropic implementation\cite{SUN19941905}
	\begin{equation}
		K_{\rm Ie}=\frac{\sqrt{\gamma_{\rm usf}o(\theta,\phi)}}{F_{12}(\theta)},
	\end{equation}
where $o(\theta,\phi)$ is a function of $\theta$ and $\phi$. $\theta$ is the angle between slip plane and crack plane, $\phi$ is the angle between the slip direction and a vector lying on slip plane and perpendicular to the crack-front direction, and $\gamma_{\rm usf}$ is the unstable stacking fault energy of the slip plane.
The computation details of $o(\theta,\phi)$ and $F_{12}(\theta)$ are given in \textit{Supplementary Material S2.3}.
These potential-dependent properties (elastic constants, $\gamma_{\rm s}$ and $\gamma_{\rm usf}$) used for computing $K_{\rm G}$ and $K_{\rm Ie}$ are listed in \textit{Supplementary Material S3}.
For $(100)[010]$ (crack plane/crack front) crack system, no IAP predicts pure cleavage on the original crack plane, in stark contrast with both the low-temperature experimental observations\cite{Hribernik_thesis} and the theoretical predictions ($K_{G}$ and $K_{\rm Ie}$).
Instead, all IAPs predict either cleavage on \{110\} planes (rather than \{100\} planes), or phase transition, or cleavage accompanied by phase transition.
With only two exceptions (MEAM-Asadi and MEAM-Etesami), the predictions of $(100)[011]$ are in good agreement with experiments\cite{Hribernik_thesis} and Griffith theory. 
For $(110)[001]$, no IAP is able to predict pure cleavage on \{100\} plane.
Cleavage and dislocation emission are both observed in $(110)[1\bar{1}0]$ crack system.
In summary, most classical IAPs predict cleavage for $(100)[011]$ and $(110)[001]$  crack systems while yielding conflicting results for the two remaining crack systems, thus leaving the question of what atomistic mechanisms control crack-tip behaviour of bcc iron unresolved. \\
\indent The lack of agreement between atomistic predictions and experiments of fracture modes and critical stress intensity factors motivates the development of a potential that is capable of reproducing the experimental observations. 
Machine learning (ML) potentials enable MD/MS simulation to describe a complex potential energy surface (PES) with density functional theory (DFT) accuracy, and are order of magnitudes faster than DFT.
Among the existing ML potential frameworks, Gaussian Approximation Potential (GAP) has been shown to accurately describe the complex motion of screw dislocations in bcc iron\cite{maresca2018screw} and the cleavage process in silicon\cite{gap_silicon_2018}. 
As a Gaussian process regression method, GAP can predict both the mean value and the variance of
the atomic energies\cite{gap_tutorial_2015}.
The square root of the GAP predicted variance is used as an indicator for error, which is useful to evaluate the quality of the model and practical in iterative/active learning\cite{GP0_PhysRevLett.122.225701,GP1_PhysRevB.100.014105}.\\
\indent We first consider an existing GAP for ferromagnetic bcc iron\cite{Dragoni2018_PhysRevMaterials.2.013808}, hereafter named Fe-GAP18.
The Fe-GAP18 database has been developed to study thermodynamics and defects in bcc iron\cite{gap18_db}.
The database includes stretched primitive cells, point defects, interfaces \textit{etc}.
Using the GAP predicted variance, we show (Fig. \ref{fig:old_gap}\textbf{a}) that Fe-GAP18 cannot predict the fracture behaviour of bcc iron.
In particular, Fig. \ref{fig:old_gap}\textbf{a} reports the maximum per-atom energy error of crack-tip atoms during mode-I MS simulations at multiple applied $K_{\rm I}'s$. 
Only crack propagation along $(110)[1\bar{1}0]$ is predicted within an accuracy of $\sim$10 meV/atom.
Fig. \ref{fig:old_gap}\textbf{b} shows snapshots with typical artifacts produced during the fracture process. 
We argue that Fe-GAP18 is unable to capture the fracture behaviour due to the lack of crack tip atomic environments in the DFT training dataset rather than due to an intrinsic limitation of GAP.
Therefore, here we develop a systematic strategy to improve GAP for fracture predictions, including a preliminary fracture-relevant database extension, which is subsequently supplemented with an active learning algorithm.
We show that the new database enables GAP to predict atomistic fracture mechanisms within an accuracy of 8 meV/atom at T=0K and on the order of $\sim$10 meV/atom at finite temperatures. \\
The paper is organised as follows. 
We describe the approach and the active learning algorithm used to expand the DFT database in section\textit{ Results "Preparation of the Reference Database"}. 
In section \textit{Results "Implementation of active learning"}, we implement the algorithm and show the convergence of our approach (hereafter indicated as Fe-GAP22). 
Based on Fe-GAP22, we predict the fracture behaviour of single crystal bcc iron at T=0K and at finite temperatures under high loading rate conditions in section \textit{Results "Prediction of fracture mechanisms and critical stress intensity factors ($K_{\rm Ic}$)"}. 
In section \textit{Discussion}, we summarise our findings and outline further research directions to be explored with the novel, quantum-accurate Fe-GAP22. 

\section*{Results}
\subsection*{Preparation of the Reference Database}
\subsubsection*{Preliminary Fracture-relevant Database Enrichment}

We first extend the existing DFT database of Fe-GAP18\cite{gap18_db} with primitive cells strained according to the large deformations that may occur at crack tips.
The strain states near a crack tip under plane strain and plane stress conditions are estimated from classical linear elastic crack-tip fields. 
Thus, we find that the maximum tensile strain is 0.27 at the point that is 0.1 nm away from the crack-tip under $K_{\rm I}=$1.5$\ \rm MPa\sqrt{m}$.
By analysing the strain components of the primitive cell from the original database\cite{gap18_db} comprising $6,000$ distorted primitive bcc cells, we found that $95\%$ of original data is within the strain of $0.15$ and the rest are highly stretched along one crystal direction ($2\sim 4$ times). 
We therefore create a new set of highly strained primitive cells to enrich the DFT dataset to encompass the classical linear elastic predicted strain states, referred to as DB9.
Note that the original database from Dragoni\cite{gap18_db} includes 8 subsets, referred to as DB1-DB8.
In order to ensure an unbiased sampling of atomic environments, we use the random uniform sampling algorithm \textit{SOBOL}\cite{sobol1967distribution} instead of the strain states directly obtained by classical elastic crack-tip fields (see \textit{Supplementary Material S4}).
Principal component analysis of the strain components shows that DB9 expands the original database and encompasses a more extensive set of strain states than those calculated by LEFM (see \textit{Supplementary Material S5}).
Common neighbour analysis shows that the original database and DB9 also include face-cubic centred (fcc) structures.
Besides, we include distorted hexagonal close-packed (hcp) primitive cells (referred to as DB10) to ensure that GAP is capable of predicting the hcp phase. \\
\indent The magnetic contribution to the total energy in iron plays an important role and thus cannot be neglected. 
Our DFT calculation of the primitive bcc cell shows that the magnetisation degree decreases when the volume decreases.
The primitive cell becomes completely non-magnetic when the volume reduction is larger than 37.5\%.  
We found that including magnetic states beyond ferromagnetic deteriorates the performance of GAP, leading to unrealistic predictions, such as negative surface energy.
This is because the transition between ferromagnetic and other magnetic states (e.g. non-magnetic) is not unique (i.e., not smooth and not single-valued), which gives rise to a partly spurious GAP predicted PES since magnetism is not explicitly accounted for in the current GAP implementation. 
Depending on the magnetisation parameter settings in DFT calculations, the self-consistent calculation may be stuck in a ground state of ferromagnetism that has larger energy than the non-magnetic ground state, or vice-versa.
The mixture of configurations with the same atomic environment and distinct magnetic states leads to a spurious GAP fitted PES.
Here, we make the assumption that the crack tip stays ferromagnetic, since the locally highly strained crack-tip bonds are surrounded by ferromagnetic iron atoms; for this reason, we train Fe-GAP22 on a consistent database of ferromagnetic configurations. 
Furthermore, DFT calculations used within the active learning scheme (\textit{see next Section}) show that the crack-tip atoms stay ferromagnetic (see \textit{Supplementary Material S6}).\\
\indent Bond rupture and dislocation emission from the crack tip are two fundamental crack-tip mechanisms\cite{ANDRIC_2017}. 
The DFT database should be sufficiently informed with configurations relevant to those mechanisms. 
Maresca \textit{et al.} showed that Fe-GAP18 predicts the single-hump Peierls potential and the compact core structure of screw dislocations with DFT accuracy\cite{maresca2018screw}. 
However, Fe-GAP18 does not include the separation process of atomic planes explicitly. 
Here, we add rigid separation configurations of \{100\}, \{110\}, \{111\} and \{112\} crystallographic planes for thin and thick slabs (referred to as DB11). 
The rigid separation includes two sets, i.e., ideal crystal structures and "rattled" structures. 
The rattled structures are obtained by adding a Gaussian noise $\mathcal{N}\sim[0,0.05 a_0]$ ($a_0$ lattice constant) to the three Cartesian coordinates of all atoms to explore the PES beyond the highly symmetric, minimum energy configurations.
In either case, separations are performed from 0 to 4 Å with a step of 0.4 Å. \\
\indent The detailed information about the preliminary database is listed in \textit{Supplementary Material S7.} 
As shown in Fig. \ref{fig:convergence_gap_error_0K}, from the original database (\textit{"Original"}) to preliminary database (\textit{"Iter-0"}), adding those relevant configurations considerably reduces the maximum GAP predicted per-atom energy error during the fracture simulations.
Yet, GAP based on \textit{"Iter-0"} database still cannot predict the fracture process within a target predicted energy error (see \textit{Supplementary Material S8}).

\subsubsection*{Active Learning Database}

Since the crack tip deformation fields are inhomogeneous, while DB9 and DB10 include only homogenous deformations, the \textit{"Iter-0"} GAP is not well-informed about crack tip local atomic environments.
By virtue of the localisation assumption that the atom only interacts with its neighbouring atoms within a finite cut-off radius (5 Å in this study), the crack tip atomic environment can be represented by a computable DFT cell that contains the atomistic crack tip configuration. 
To learn efficiently from the crack tip, we develop an active learning algorithm by extracting crack tips from extrapolating configurations. 
The algorithm (see \textit{Data Availability}) automatically identifies the extrapolated atomic environments with respect to a predefined criterion and constructs a periodic crack cell that is computable with plane-wave DFT while minimising spurious surface effects. 
The GAP predicted variance per-atom is a measure of the extrapolating degree from the existing DFT dataset, which is used to assess the accuracy of the new configurations during the simulation.
The square root of GAP predicted variance per-atom is compared with a predefined criterion to determine the extrapolation.
The predefined criterion is set to 10 meV/atom in this study.
 \\
\begin{figure}[H]
	\centering
	\includegraphics[scale=0.5]{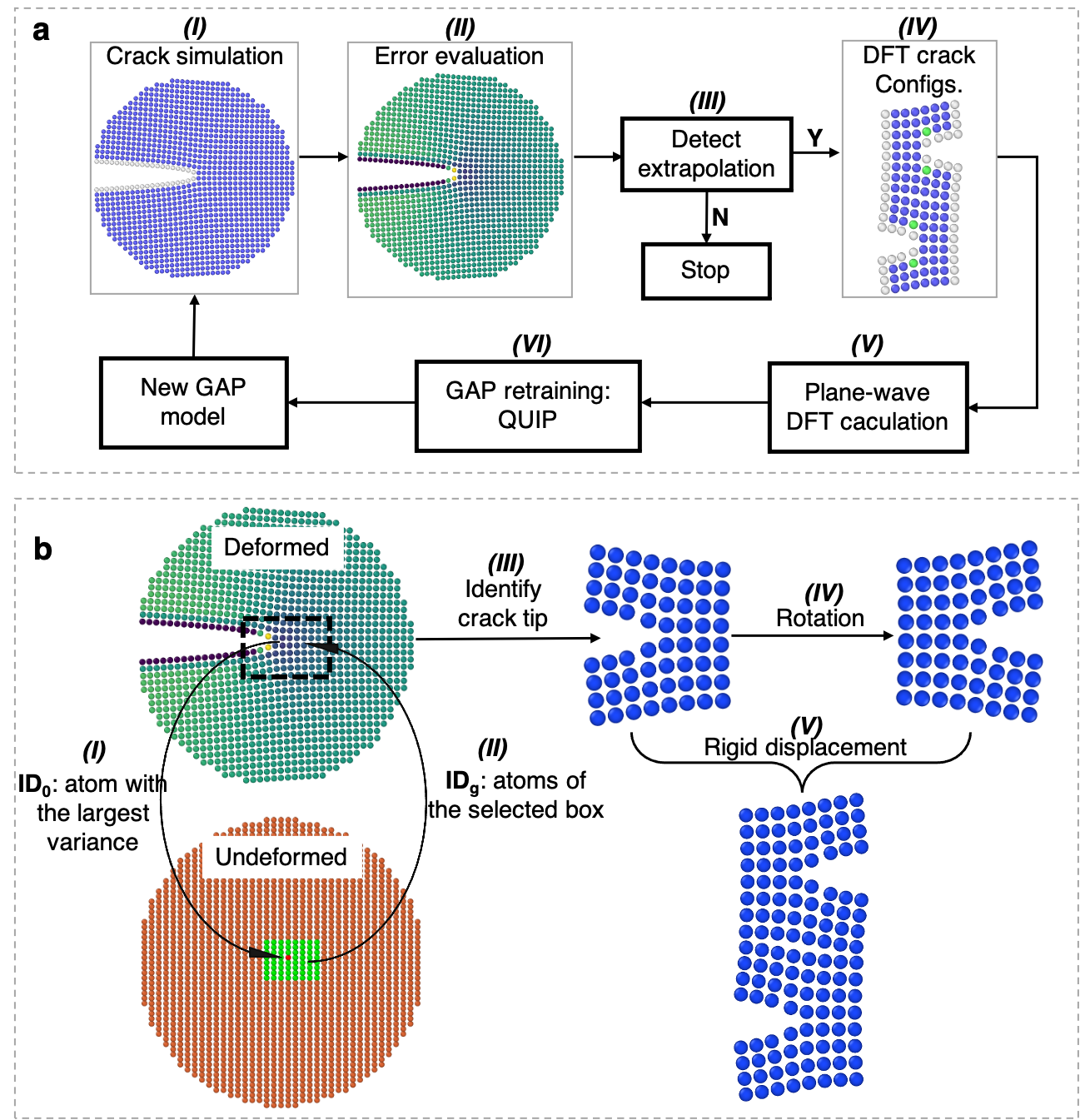}
	\caption{
	\textbf{(a)} Schematic workflow of the active learning algorithm.
		\textit{(I)} Fracture simulation by $K$-test at T=0K. 
		\textit{(II)} Evaluation of the GAP predicted per atom energy error for each frame generated during $K$-test . 
		\textit{(III)} Detection of extrapolation by comparing the maximum GAP predicted error per atom of each configuration with the predefined criterion.
		 \textit{(IV)} Construction of the crack-tip cell that is computable by DFT.
		 \textit{(V)} DFT calculation of the crack-tip cell. 
		 \textit{(VI)} Refitting GAP. 
		 Steps \textit{(I)$\sim$(VI)} are repeated until convergence is achieved. 
	\textbf{(b)} Construction of the crack-tip DFT cell. 
		\textit{(I)} Identification of the atom with the largest error ($\rm{ID}_0$)  in the deformed configuration. 
		\textit{(II)} Selection of a group of atoms ($\rm{ID}_g$) in the square centred on $\rm{ID}_0$ in the undeformed state.
	 	\textit{(III)} Identification of the crack tip in group $\rm{ID}_g$ in the deformed state. 
	 	\textit{(IV)} Duplication and rotation of the selected region.
	  	\textit{(V)} Generation of DFT cell containing two crack tips by merging the rotated replica and the original ones.
  	}
	\label{fig:schema_active}
	\vspace{-7.5mm}
\end{figure}
\begin{table}[H]
	\centering
	\caption{Summary of active learning algorithm. Roman numbers correspond to the ones shown in Fig. \ref{fig:schema_active}\textbf{a}. }
	\begin{tabular}{|c|c|c|} 
		\hline
		Step & Function &  Code \\ [0.5ex] 
		\hline
		\textit{\textbf{\MakeUppercase{\romannumeral 1}}} & $K$-test & LAMMPS\cite{LAMMPS} \\ 
		\textit{\textbf{\MakeUppercase{\romannumeral 2}}} & Evaluation of the GAP predicted variance & QUIP\cite{csanyi2007expressive} \\
		\textit{\textbf{\MakeUppercase{\romannumeral 3}}} & Detection of extrapolation & Python \\
		\textit{\textbf{\MakeUppercase{\romannumeral 4}}} & Construction of the DFT cell & Python \\
		\textit{\textbf{\MakeUppercase{\romannumeral 5}}} & DFT calculation & Quantum Espresso\cite{Giannozzi_2009} \\ 
		\textit{\textbf{\MakeUppercase{\romannumeral 6}}} & Fitting GAP model & QUIP \\
		\hline
	\end{tabular}
	\label{table:1}
	\vspace*{-5mm}
\end{table}
Fig. \ref{fig:schema_active}\textbf{a} shows the workflow, schematically indicating the main steps of the active learning algorithm. 
We first run a $K$-test and evaluate the GAP predicted variance for all frames dumped during the fracture simulation. 
Then, the GAP predicted per-atom energy error is compared with the predefined extrapolation criterion. 
A periodic cell is subsequently constructed by extracting the atomic configuration from the extrapolating frame.
DFT calculations of the periodic crack-tip cell are performed, and the new DFT data are added to the training database.
Next, GAP is retrained with the new DFT database.
We repeat this process until convergence is achieved, i.e., the maximum GAP predicted per-atom energy error is less than the predefined criterion during the entire fracture process.
A summary of the software/code used in the active learning scheme is listed in Table \ref{table:1}.\\
\indent One of the main challenges is the construction of an appropriate DFT cell that contains the crack tips in step \textit{(IV)} of Fig. \ref{fig:schema_active}\textbf{a}. 
It is necessary to prevent spurious boundaries to avoid learning irrelevant/artificial atomic environments. 
The active learning has been applied to study the screw dislocation in bcc tungsten, in which the effective extrapolative configuration is constructed by taking advantage of symmetry\cite{hodapp2020operando}.
The idea here is to symmetrise the crack tip along the crack plane normal to construct an approximate periodic cell.
The construction of the DFT crack configuration is illustrated in Fig. \ref{fig:schema_active}\textbf{b}. 
First we identify the atom ($\rm{ID}_0$) with the largest GAP predicted error at the crack tip in the deformed configuration. 
Then, a group of atoms ($\rm{ID}_g$) in a square centred on $\rm{ID}_0$ is selected in the undeformed configuration;
once identified, these atoms form a crack tip configuration in the deformed state. 
Another crack tip configuration is created by duplicating the original one and rotating it by $180^{\circ}$. 
We construct the periodic DFT configuration by aligning the rotated replica and the original one by a rigid displacement. 
The periodic DFT configuration is surface-free and preserves the local atomic environment of the atom with the largest GAP predicted error.
Such construction allows both efficient DFT calculations and learning speed, as illustrated in the following section.

\subsection*{Implementation of active learning}

\begin{figure}[H]
	\centering
	\includegraphics[trim=0 30 0 70, clip, scale=0.42]{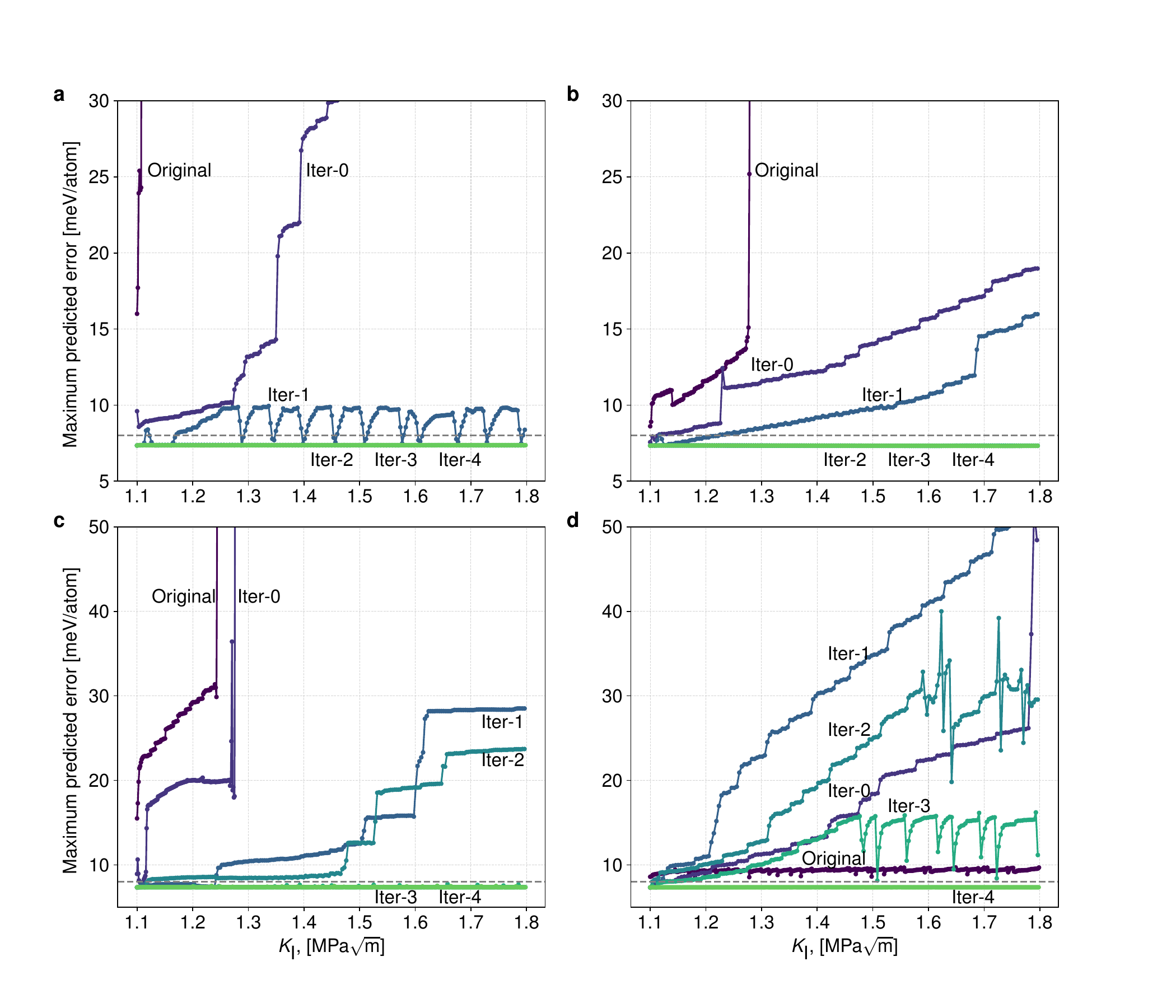}
	\caption{Maximum GAP predicted energy error per atom as functions of $K$ for four crack systems: (a)$(100)[011]$, (b)$(110)[001]$, (c)$(100)[010]$ and (d)$(110)[1\bar{1}0]$. The dashed line indicates the 8 meV/atom. \textit{Original} is the GAP trained on the original database\cite{gap18_db} (see Fig. \ref{fig:old_gap} for the value of error). \textit{Iter-0} is the GAP trained only with preliminary database.}
	\label{fig:convergence_gap_error_0K}
	\vspace*{-3mm}
\end{figure}

We implemented the active learning algorithm with the original SOAP\cite{descriptors_Albert_2013} and an optimised SOAP descriptor\cite{turboSOAP_PhysRevB.100.024112} (referred to as "Turbo SOAP"). 
The presentation here focuses on active learning results with Turbo SOAP because this leads to higher computing speed and faster convergence.
The results based on the original SOAP are discussed in \textit{Supplementary Material S9}. 
We applied the algorithm to two crack systems in parallel, i.e., $(100)[011]$ and $(110)[001]$, such that GAP can learn the local environments from both $\{100\}$ and $\{110\}$ crack planes at the same time.
Two sets of small crack tip regions are used, i.e., configurations extracted from fracture simulations with and without Gaussian noise ($\mathcal{N}\sim[0,0.05a_0]$) added to the three Cartesian coordinates. 
Fig. \ref{fig:convergence_gap_error_0K} shows the convergence of the maximum GAP predicted per-atom energy error at multiple $K_{\rm I}$'s with respect to active learning iterations. 
The GAP predicted energy error is plotted as a function of $K_{\rm I}$ during a mode-I $K$-test at T=0K.  \\
\indent As shown in Fig. \ref{fig:convergence_gap_error_0K}\textbf{a} and \textbf{b}, the maximum GAP predicted error converges to 8 meV/atom for crack systems $(100)[011]$ and $(110)[001]$ after two iterations of active learning.  
The GAP predicted error of $(100)[010]$ converges to 8 meV/atom at the third iteration (Fig. \ref{fig:convergence_gap_error_0K}\textbf{c}) while $(100)[011]$ and $(110)[001]$ crack systems remain of the same accuracy.
As shown in Fig. \ref{fig:convergence_gap_error_0K}\textbf{d}, the GAP predicted error of $(110)[1\bar{1}0]$ increases after two iterations (\textit{Iter-0} and \textit{Iter-1}) and then decreases but does not converge to 8 meV/atom after another two iterations (\textit{Iter-1} and \textit{Iter-2}). 
This is because the original GAP is able to extrapolate the atomic environments that are close to crack tip of $(110)[1\bar{1}0]$.
However, the extrapolation is unstable and is easily deteriorated by adding more data, which means that the prediction is not converged yet. 
We further added a crack tip of $(110)[1\bar{1}0]$ for one additional iteration of training and converged the GAP predicted error to 8 meV/atom (Fig. \ref{fig:convergence_gap_error_0K}\textbf{d}).

\subsection*{Prediction of fracture mechanisms and critical stress intensity factor ($K_{\rm Ic}$)}

\begin{figure}[H]
	\centering
	\includegraphics[trim=0 0 0 0, clip, scale=0.36]{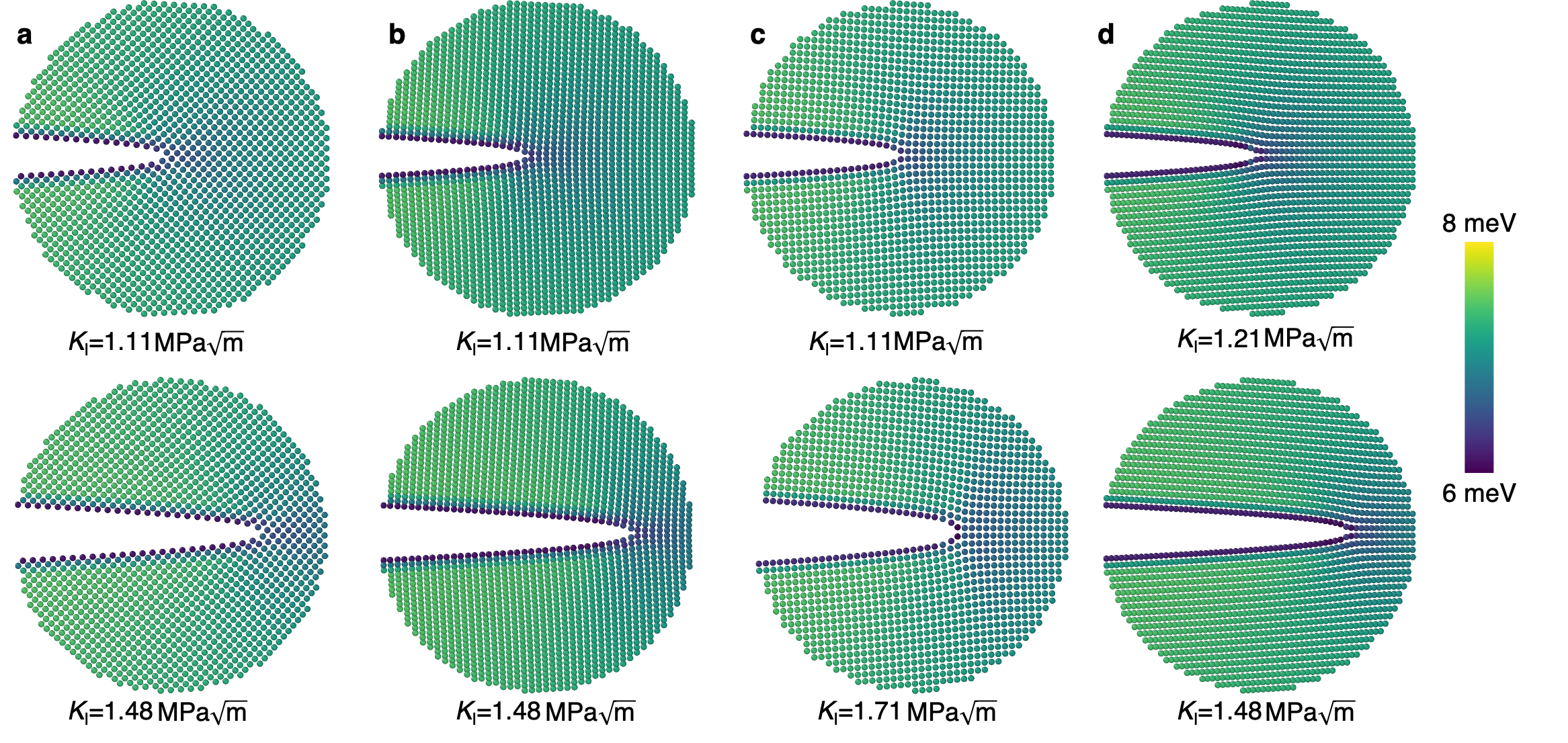}
	\caption{Atomic snapshots showing the fracture mechanism at T=0K predicted by Fe-GAP22. 
		\textbf{(a)}$(100)[010]$, \textbf{(b)}$(100)[011]$, \textbf{(c)}$(110)[001]$, and \textbf{(d)} $(110)[1\bar{1}0]$. 
		The configuration is taken at $K_{\rm I}$ given below each snapshot. 
		Atoms are coloured by the predicted energy error per atom. 
	}
	\label{fig:snapshots_0K}
\end{figure}

\begin{figure}[H]
	\centering
	\includegraphics[trim=0 10 0 40, clip, scale=0.4]{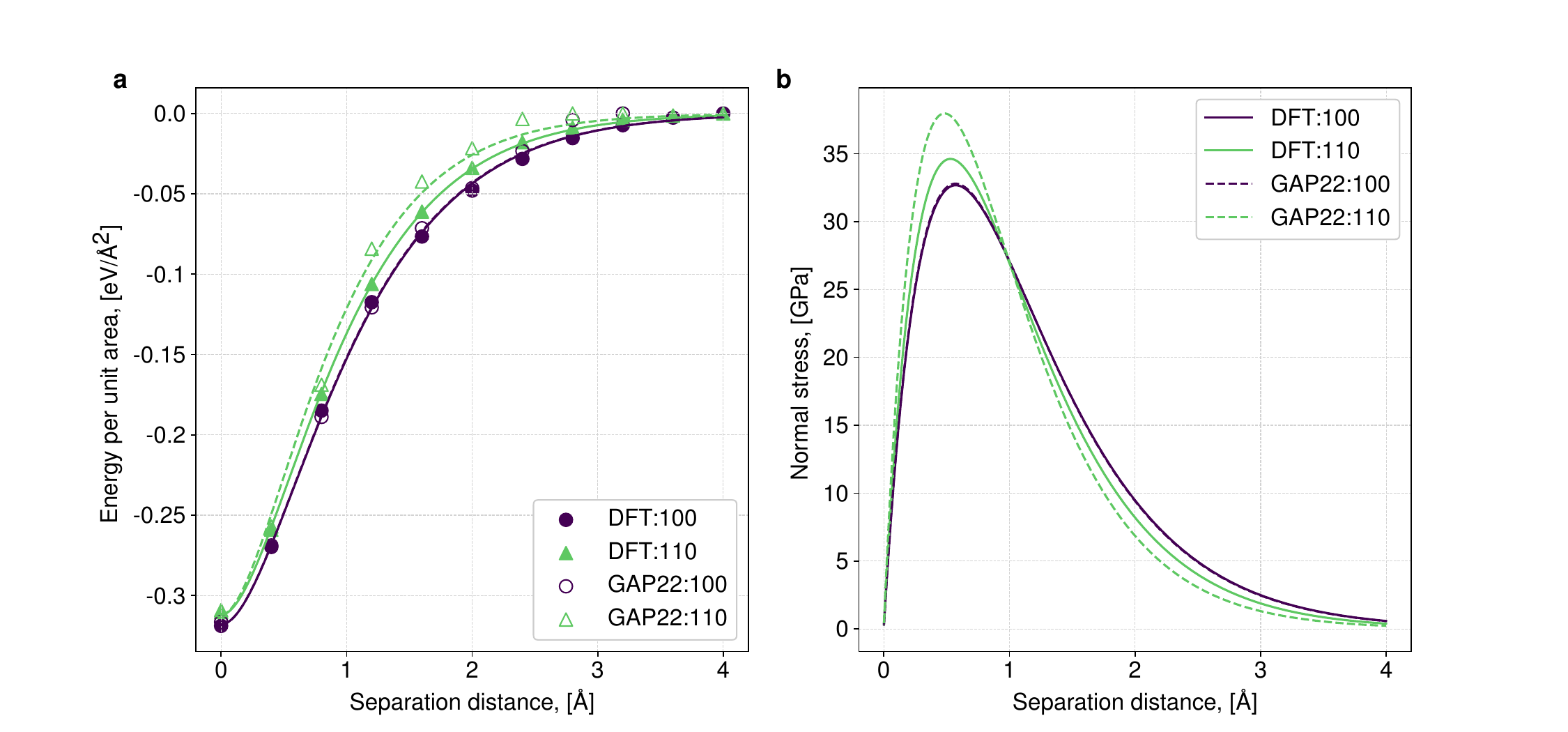}
	\caption{\textbf{(a)} Universal binding energy relation (UBER) for $\{100\}$ and $\{110\}$ plane\cite{rydberg_PhysRevB.28.1835}. Filled and unfilled symbols represent DFT and Fe-GAP22 predictions respectively. Circle and triangle symbols indicate $\{100\}$ and $\{110\}$ planes respectively. Solid and dash lines are DFT and Fe-GAP22 results fitted to UBER equation.
	\textbf{(b)} Normal stress on $\{100\}$ and $\{110\}$ planes by taking derivative of fitted curves from \textbf{(a)}. 
		}
	\label{fig:uber}
\end{figure}

\begin{figure}[H]
	\centering
	\includegraphics[trim=0 20 0 50, clip, scale=0.4]{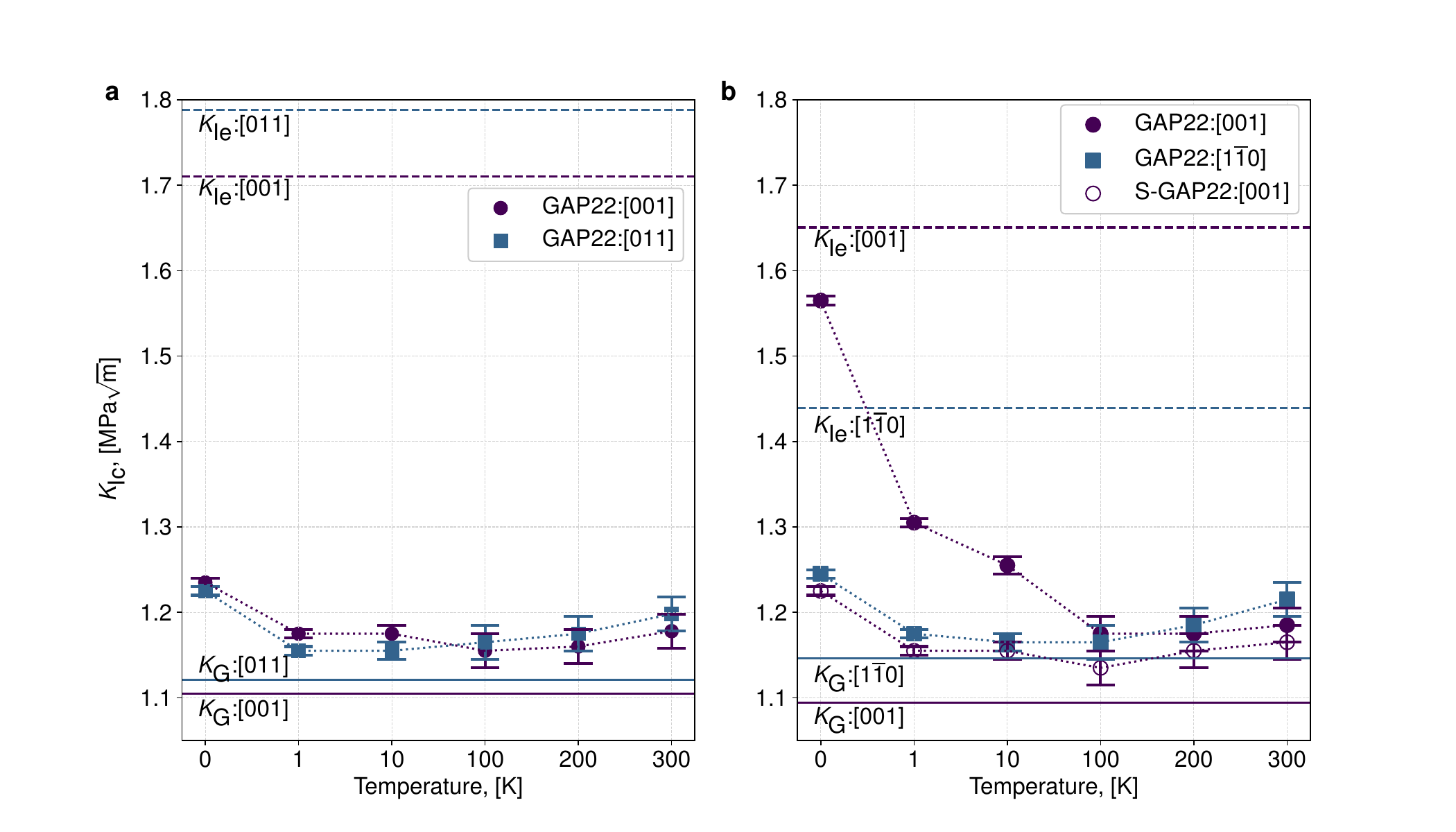}
	\caption{Critical $K_{\rm I}$ predicted by Fe-GAP22 at different temperatures (T=0K$\sim$300K). 
		\textbf{(a)}(100) and \textbf{(b)}(110) crack plane. 
		The solid and dashed line indicate critical $K$ predicted by Griffith ($K_{\rm G}$) and Rice ($K_{\rm Ie}$) theories, respectively. 
		Fe-S-GAP22 is the GAP trained on the active learning database developed based on original SOAP.
		MD simulations at finite temperatures are performed 5 times to get the average of the predicted $K_{\rm Ic}$. 
	}
	\label{fig:finite_temp_K}
\end{figure}

We predicted the lattice constant, elastic constants and surface energies as a benchmark of the newly developed Fe-GAP22 and found similar accuracy as Fe-GAP18 (\textit{see Supplementary Material S10}).
Subsequently, we performed fracture simulation of single-crystal ferromagnetic iron at T=0K and finite temperatures (T=1, 10, 100, 200, 300K) under high loading rate ($10^9\ \rm MPa\sqrt{m}s^{-1}$).
The fracture mechanism at T=0K is found to be cleavage on the original plane in all considered cases, and the GAP predicted per-atom energy error remains below 8 meV/atom during the entire cleavage process (Fig. \ref{fig:snapshots_0K}).
Neither phase transformation nor $\{110\}$ planar faults appear according to Fe-GAP22 predictions. \\
\indent The cleavage is always taking place on the plane that has the maximum normal stress. 
To evaluate the normal stress, we performed rigid body separation of $\{100\}$ and $\{110\}$ planes by using DFT and Fe-GAP22 to calculate the universal binding energy relation (UBER) (Fig. \ref{fig:uber}\textbf{a}).
Fe-GAP22 predicted UBER of $\{100\}$ plane perfectly overlaps with DFT ones, while the Fe-GAP22 predicted UBER of $\{110\}$ plane slightly deviates from DFT calculations.
We further computed the normal stress during the separation (Fig. \ref{fig:uber}\textbf{b}) by taking the derivative of the fitted curve in Fig. \ref{fig:uber}\textbf{a}. 
The minimum stress required to separate $\{100\}$ and $\{110\}$ surfaces predicted by DFT are 32.68 and 34.62 GPa, respectively, and Fe-GAP22 predictions are 0.3\% and 9.7\% larger than DFT results.
For all crack systems considered here, the deviation of the crack from the original plane to $\{100\}$ or $\{110\}$ is not likely to occur because it requires a lower cohesive strength than on the original plane. 
The Fe-GAP22 prediction of cohesive strength confirms that the cleavage will take place on the original crack plane, which is also consistent with anisotropic LEFM calculations, i.e., the largest normal stress exists on the original plane under pure mode-I loading (see \textit{Supplementary Material S11}). \\
 At finite temperatures, cleavage is consistently taking place on the original plane for all crack systems, which shows that cleavage is the controlling mechanism of fracture in iron at high loading rates when considering finite temperatures (T$\leq$300K).  
At T=300K, the GAP predicted error raises to $\sim$30 meV/atom for $(110)[1\bar{1}0]$ and $\sim$10 meV/atom for the other three crack systems (see \textit{Supplementary Material S12}).
Five independent $K$-tests are performed at each temperature for the statistic prediction of $K_{\rm Ic}$ at finite temperatures, as indicated by the error bar shown in Fig. \ref{fig:finite_temp_K}. 
Here we propose a criterion to determine $K_{\rm Ic}$ for cleavage from atomistic simulation results based on the traction-separation curve. 
The first pair of atoms at the crack tip can still interact with each other after the initial debonding process until the entire separation is achieved, i.e. the distance between them reaches the cutoff distance in atomistic simulations.
Therefore, $K_{\rm Ic}$ is defined as the point where the crack tip bond is separated to a cutoff distance and will remain open upon further increasing $K_{\rm I}$.
We also considered the GAP trained on the database developed based on the original SOAP descriptor (see \textit{Supplementary Material S9}), referred to as Fe-S-GAP22 in Fig. \ref{fig:finite_temp_K}. 
Inspection of the atomic snapshots reveals that the fracture mechanism for all crack systems remains cleavage up to T=300K as predicted by Fe-S-GAP22, which demonstrates that the predicted cleavage mechanism is independent of the specific crack-tip converged DFT database.\\
\vspace*{-2mm}
Fe-GAP22 predicted $K_{\rm Ic}$ for all crack systems decreases from 0K to 1K and converges from T=1K to 200K for all crack systems except for $(110)[001]$ (Fig. \ref{fig:finite_temp_K}). 
The decrease from 0K to 1K is around $0.05 \rm MPa\sqrt{m}$, which is an accuracy limit rather than lattice trapping effects and is due to negligible barriers.
Those minute barriers might not be overcome by geometry optimisation/static calculations.
Instead, low temperature MD helps to overcome small barriers and move away from saddle points/symmetric configurations.
Fe-GAP22 prediction of $(110)[001]$ significantly decreases from T=0K to 10K and converges at T>100K, which is a consequence of the rough GAP predicted PES.
Fig. \ref{fig:finite_temp_K}\textbf{b} shows the Fe-S-GAP22 predicted $K_{\rm Ic}$ for $(110)[001]$ (see \textit{Supplementary Material S13} for prediction of the other 3 crack systems).
The predicted $K_{\rm Ic}$'s of Fe-GAP22 and Fe-S-GAP22 converge to the same value within a 0.004 $\rm MPa\sqrt{m}$ error bound.
All predictions at T>10K are within $10\%$ of Griffith prediction ($K_{\rm G}$), indicating mild lattice trapping effects\cite{curtin1990lattice}.
At high temperatures (T>100K), the surfaces near the crack tip can be closed due to thermal fluctuations, which leads to a slight increase of $K_{\rm Ic}$ at T=200K and 300K for all crack systems. 

\section*{Discussion}

Our analysis unveils that the T=0K fracture mechanism for \{100\} and \{110\} crack planes is cleavage on the pre-existing crack planes. 
There are no fracture experiments performed close to T=0K, thus qualitative comparison is possible only with the work of Ref.\cite{Hribernik_thesis}, which is performed at T=77K and finite loading rates that are inaccessible to direct MD simulations.
The prediction of Fe-GAP22 qualitatively agrees with experiments on $\{100\}$ crack plane, i.e., cleavage on the original crack plane. 
This is a considerable step forward with respect to all classical potentials considered by M\"{o}ller\cite{Moller_2014} and in the introduction of the present paper, that, in contrast with Fe-GAP22, cannot predict pure cleavage on $\{100\}$ plane for $(100)[010]$ crack system.
Furthermore, Fe-GAP22 is free from commonly observed artifacts of classical potentials, such as the occurrence of $\{110\}$ planar faults\cite{moller2018110}. 
However, in the case of $\{110\}$ crack planes, Fe-GAP22 also predicts cleavage on the original plane while experiments show crack kinking into $\{100\}$ planes\cite{Hribernik_thesis}.
Our calculation of UBER by Fe-GAP22 and anisotropic LEFM analysis shows that, under $K_{\rm I}$ loading and in the absence of other defects close to the crack tip, cleavage should take place on the original crack plane for both $\{100\}$ and $\{110\}$. 
The experiments, instead, show a semi-brittle fracture behaviour of bcc iron because cleavage is accompanied by extensive dislocation activity\cite{Hribernik_thesis}.\\
\indent Therefore, in order to perform quantitative connection to experiments, the temperature and loading-rate dependent competition between cleavage and dislocation emission must be assessed, e.g. by performing Nudged Elastic Band (NEB) calculations and using transition state theory\cite{MAK2021104389,zhu2004atomistic}. 
Also, the interaction between pre-existing defects (e.g. nanovoids, dislocations, \textit{etc}) and cracks is expected to change the stress field around the crack tip and hence affect the crack-tip response.
In the present atomistic fracture simulations however, the specimen is defect-free (except for the sharp crack),
and the boundary conditions are those of a half-infinite crystal in a 2D setup. 
Therefore, quantitative predictions of fracture require a multiscale approach.
Despite the fact that a quantitative comparison of $K_{\rm Ic}$ is not possible, we extrapolate the experimental $K_{\rm Ic}$ linearly to T=0K.
The experimental $K_{\rm Ic}$ for \{100\} and \{110\} crack planes are $\sim$4 and $7\sim$12 $\rm MPa\sqrt{m}$, respectively\cite{Hribernik_thesis}. 
$K_{\rm Ic}$ for \{110\} crack plane is $2\sim3$ times larger than \{100\}, suggesting pronounced thermally-activated dislocation plasticity at finite temperatures on \{110\} crack plane. 
Such a significant energy dissipation is expected to account for the crack deviation from \{110\} to \{100\}crack plane.
Thus, large-scale simulations in which rate and temperature dependence is considered along with pre-existing defects should be performed to enable prediction of the experimental $K_{\rm Ic}$.\\
\indent Griffith\cite{Griffith} ($K_{\rm G}$) and Rice\cite{RICE1992239} ($K_{\rm Ie}$) theories are used to predict cleavage and dislocation emission from the crack tip.
Since cleavage is the controlling fracture mechanism in bcc iron at T=0K, we compare the Fe-GAP22 predicted $K_{\rm Ic}$ with Griffith theory $K_{\rm G}$.
Because $K_{\rm G}$ only accounts for the energy of newly created surface, it serves as a lower bound of $K_{\rm Ic}$.
Thus, the excess part of Fe-GAP22 predicted $K_{\rm Ic}$ compared with $K_{\rm G}$ could be used as an indicator of lattice trapping effects.
For all three crack systems except for $(110)[1\overline{1}0]$, the predicted $K_{\rm Ic}$ is around $10\%$ larger than Griffith criterion, which is also found in bcc high entropy alloys\cite{MAK2021104389}. 
$(110)[1\overline{1}0]$ is $5\%$ larger than $K_{\rm G}$, indicating less lattice trapping effects.\\
\indent We obtained a Fe GAP that is able to predict atomistic fracture in iron with an accuracy of 8 meV/atom based on an existing database, showing that GAP can be improved for fracture study. 
Our strategy can be generalised to other materials and other ML frameworks, providing a systematic way to improve machine learning potentials for studying fracture behaviour.
The current approach is based on two essential ingredients, i.e., fracture-relevant configurations and an active learning database.
Based on the fracture-relevant database, our active learning approach produced convergence to predicted energy error of 8 meV/atom in 4 iterations. 
We found that crack propagates on the original crack plane for $(100)[010]$, $(100)[011]$, $(110)[001]$ and $(110)[1\overline{1}0]$ crack system.
Based on the convergence of Fe-GAP22 predicted $K_{\rm Ic}$ with respect to temperatures, we conclude that the increase of $K_{\rm Ic}$ at T=0K is caused by minute barriers.
The predicted $K_{\rm Ic}$ at T=0K is converged with respect to local symmetry breaking, which we have studied by performing high loading rate mode-I tests from T=10K up to T=300K.
Our results confirm the T=0K atomistic fracture mechanism in ferromagnetic iron under mode-I loading, settling the inconsistency of crack-tip mechanisms based on classical potentials. \\
\indent We conclude by pointing out that Fe-GAP22 can describe cracks and screw dislocations with quantum accuracy\cite{maresca2018screw}.
The Fe-GAP22 is currently being applied to calculate the energy barriers for the fracture-related processes with NEB.
Based on transition state theory, the competition of atomistic crack-tip mechanisms (cleavage and dislocation emission) can be connected to microscale behaviour to assess the brittle to ductile transition\cite{MAK2021104389} at finite temperatures and finite loading rates.
Mobility laws for microscale simulations can thus be developed by using Fe-GAP22.

\section*{Methods}

\subsection*{DFT calculation}

All DFT calculations were performed in collinear spin-polarised plane wave method implemented within QUANTUM ESPRESSO\cite{Giannozzi_2009}. 
An ultrasoft GGA PBE pseudopotential from $0.021 pslibrary$ is employed\cite{PBE}. 
The kinetic energy cutoff for wavefunctions and charge density are set to 90 Ry and 1080 Ry, respectively. 
The Brillouin Zone is integrated with a Monkhorst-Pack grid and a Marzari-Vanderbilt smearing scheme at an effective temperature of 0.01 Ry\cite{MV-smearing_PhysRevLett.82.3296}. 
The starting value of magnetisation is set to 0.34.  
An adaptive k-mesh approach is used to calculate primitive cells (DB9 and DB10), ensuring that the $k$-spacing is smaller than 0.015 Å$^{-1}$. 
For other calculations, the simulation cell size is chosen to satisfy the condition that $k$ spacing is smaller than 0.03 Å$^{-1}$ while only the $\Gamma$ point is used along the separation direction. 
A benchmark study compared with Dragoni's database is conducted to ensure all the parameter settings can reach the convergence of 1 meV/atom, 0.01 eV/Å and 0.01 GPa for energies, forces and stresses, respectively. 

\subsection*{GAP training}
A parallel version (1645177290) of the open-source package QUantum mechanics and Interatomic Potentials (QUIP) is used to fit the GAP model\cite{csanyi2007expressive}. 
We use three descriptors to encode the local atomic environment, i.e., one distance-based two-body interaction and two many-body turbo-SOAP descriptors.   
Two turbo-SOAP descriptors are used to describe the inner and outer atomic environment within the cutoff radius of 3.5 Å and 5 Å, respectively. 
Dot product kernel is used for both turbo-SOAP descriptors, and the CUR matrix decomposition procedure is applied to find the optimal representative local atomic environments.
The cutoff smoothing distance is set to 1 Å. 
The number of radial and angular basis for turbo-SOAP are set to 8.
The training command line is available in \textit{Data Availability}.

\subsection*{MS/MD simulation setup}
We use a cylinder-shaped half crack setup (see schematic plot in \textit{Supplementary Material S14}), where $x$, $y$ and $z$ are aligned with the crack propagation direction, crack plane normal and crack front, respectively. 
The radius of the model in $x$-$y$ plane is set to 150Å, which satisfies the requirement that the fracture process zone is much smaller than the simulation cell, as proved in Ref. \cite{Moller_2014,Andric_2018}. 
We consider plane strain conditions in a 2D setting by imposing periodic boundary conditions along the crack front direction. \\
\indent We used a $K$-test loading scheme that allows controlling the fracture process by directly increasing the stress intensity factor $K_{\rm I}$. 
$K$-test implements a displacement-controlled loading scheme, which makes use of anisotropic LEFM. 
The displacement field of a half crack in an infinite anisotropic medium can be derived via Lekhnitskii’s formalism\cite{Ting_1996}, as summarised in \textit{Supplementary Material S2.1}.
The asymptotic stress field near the crack tip can be uniquely characterised by the stress intensity factor $K_{\rm I}$, which enables a single-parameter controlled loading scheme\cite{Andric_2018}.
The $K_{\rm I}$ displacement field can then be applied to the boundary of the simulation cell directly. 
During $K$-test, $K_{\rm I}$ is increased with a step of $\Delta K_{\rm I}=0.01 \rm MPa\sqrt{m}$. 
The displacement is applied incrementally at each step, and the system is equilibrated under the constraining of fixed boundary atoms. 
We use a combination of conjugate gradient (CG) and FIRE minimizers\cite{fire_bitzek2006structural} with a force tolerance of $10^{-9}$ eV/Å and $10^{-3}$ eV/Å, respectively.
For MD, we use the Nosé-Hoover thermostat with a timestep of 0.001 ps and the system is equilibrated for 10 ps at each incremental step.\\
\indent To preserve an atomistically sharp crack, we neither screen the interaction between free crack surfaces nor delete any atoms (so-called screening and blunting\cite{Andric_2018}). 
Instead, we start with a $K_{\rm init}$ that maintains the current crack tip position.
However, it should be noted that finding $K_{\rm init}$ may require trial-and-error. 
All molecular statics/dynamics calculations in this work are performed by using LAMMPS\cite{LAMMPS}.
The atomic configurations are visualised in OVITO\cite{ovito}.

\section*{Data Availability}

The training command line and extended DFT database are available on Materials Cloud \url{https://archive.materialscloud.org/record/2022.102}.
The active learning algorithm is available on our Github page \url{https://github.com/leiapple/Fe-GAP22-AL}.

\bibliography{sample}

\noindent\textbf{Acknowledgements}\\
This work made use of the Dutch national e-infrastructure with the support of the SURF Cooperative using grant no. EINF-2393 and EINF-3104. 
We thank the Center for Information Technology of the University of Groningen (UG) for their support and for providing access to the Peregrine high performance computing cluster. 
LZ would like to thank Predrag Andric for LAMMPS implementation of fracture simulation and Miguel Caro for the implementation of Turbo SOAP descriptor. 
FM acknowledges the support from the start-up grant from the Faculty of Science and Engineering at the University of Groningen. \\

\noindent\textbf{Author contributions}\\
All authors designed the research together. L.Z. and F.M. developed the active learning framework. L.Z. implemented the framework and performed all the DFT, MS and MD calculations. All authors analysed the data, discussed the results, wrote the manuscript together and contributed to the discussions and revisions of the paper.\\

\noindent\textbf{Competing interests}\\
The authors declare no competing interests.\\

\noindent\textbf{Supplementary information}\\
Please see the Supplementary Material for supporting information of the findings in the manuscript. 

\end{document}


\setstretch{1.5}
\maketitle

\section*{S1: A brief summary of existing atomistic studies on fracture in bcc iron}

Extensive MD simulations have been performed to characterize crack-tip behaviour in single crystal bcc iron throughout the last $\sim$40 years \cite{deCelis_1983,Guo_2007_acta,Guo_2006,Guo_2007_msea,cao_wang_2006,Wang_2021,Meiser_2020,Cheung1994AMS,Gordon2007CrackTipDM,Moller_2014}.
An edge-crack geometry in an half-infinite medium under plane strain mode-I loading conditions is mostly investigated and multiple deformation mechanisms are found at the crack tip. 
deCelis \textit{et al.}\cite{deCelis_1983} and Kohlhoff \textit{et al.}\cite{Kohlhoff_1991} observed cleavage on $\{100\}$ crack plane for $(100)[010]$ and $(100)[011]$ crack systems.
Guo \textit{et al.}\cite{Guo_2007_acta,Guo_2007_msea} found that twinning and recrystallization are the main deformation mechanisms at T=5K.
Cao \textit{et al.}\cite{cao_wang_2006} pointed out that twinning and cleavage processes cooperate at low temperatures.
Wang \textit{et al.}\cite{Wang_2021} observed deformation-induced martensitic transformations (DIMTs) in the $(010)[100]$ crack system. 
The existence of DIMTs at low temperatures is still under debate since this process is energetically unfavourable at T=0K, according to Ref.\cite{Meiser_2020}.
By studying the temperature effects, Cheung \textit{et al.}\cite{Cheung1994AMS} found that the brittle to ductile transition temperature (BDTT) in iron is between T=200K and 300K for various crack-tip geometries under mode-I loading.
Gordon \textit{et al} investigated the influence of interatomic potentials (IAPs) on crack-tip response under mode-I loading\cite{Gordon2007CrackTipDM} .
The predicted $K_{\rm Ic}$ is compared with Griffith\cite{Griffith} and Rice\cite{RICE1992239} theories ($K_{\rm G}$ and $K_{\rm Ie}$).
Gordon \textit{et al.}\cite{Gordon2007CrackTipDM} found that crack tip exhibited qualitatively different responses among four different embedded atom method (EAM) models.
M\"{o}ller and Bitzek\cite{Moller_2014} conducted a similar study on the influence of eight EAM potentials on the behaviour of atomistic cracks in single iron and concluded that no EAM potential is able to fully reproduce the experimental fracture behaviour in all crack systems of bcc iron.
The existing approaches are summarized in Table \ref{tab:summary_atomistic}. \\
\indent In our work, we consider the edge crack geometry, which enables the simulation of a $K$-dominant region that is larger than the fracture process zone and much smaller than the crack size. 
This cannot be achieved with atomistic simulations of centered and/or penny-shaped cracks\cite{Andric_2018}.
Moreover, we only consider \{100\} and \{110\} crack planes since these are the ones commonly observed in experiments of bcc iron.

\begin{table}[h]
	\begin{center}
		\caption{Method summary of existing atomistic studies.}
		\label{tab:summary_atomistic}
		\begin{tabular}{cc} 
			\hline
		    \hline
			Specimen geometry    & Centered, Penny, Edge\\ 
			\hline
			crack plane/crack front        & \{100\}\{110\}\{111\},$\langle$100$\rangle$$\langle$110$\rangle$$\langle$111$\rangle$$\langle$112$\rangle$\\ 
			\hline 
			Boundary conditions       & LEFM, homogeneous strain field  \\ 
			\hline
			Temperatures   & 0K to 300K \\
			\hline
			\hline
		\end{tabular} 
	\end{center}
\end{table}
\section*{S2: Anisotropic Linear Elastic Fracture Mechanics}

\subsection*{S2.1: Classical anisotropic linear elastic crack-tip fields}

The displacement and stress fields of a half crack in an infinite anisotropic medium can be derived via Lekhnitskii’s formulation\cite{Ting_1996}.
The governing equation of the plane strain problems in an anisotropic linear elastic medium is
\begin{equation}\label{eq:1}
	\begin{split}
		b_{11}\frac{\partial^4 U}{\partial y^4}+b_{22}\frac{\partial^4 U}{\partial x^4}+(2b_{12}+b_{66})\frac{\partial^4 U}{\partial x^2 \partial y^2}-2b_{16}\frac{\partial^4 U}{\partial x \partial y^3}-2b_{26}\frac{\partial^4 U}{\partial x^3 \partial y}=0,
	\end{split}
\end{equation}
where $U$ is the \textit{Airy stress function} and $b_{ij}$ are determined by the compliance $S_{ij}$\cite{sih1965cracks,lieberman1956simplified}
\begin{equation}\label{eq:2}
	\begin{split}
		&b_{11}=\frac{S_{11}S_{33}-S_{13}^2}{S_{33}}, \
		b_{12}=b_{21}=\frac{S_{12}S_{33}-S_{13}S_{23}}{S_{33}},\\ &b_{22}=\frac{S_{22}S_{33}-S_{23}^2}{S_{33}}, \
		b_{16}=b_{61}=\frac{S_{16}S_{33}-S_{13}S_{36}}{S_{33}},\\
		&b_{66}=\frac{S_{66}S_{33}-S_{36}^2}{S_{33}},\
		  b_{26}=b_{62}=\frac{S_{26}S_{33}-S_{23}S_{36}}{S_{33}}.
	\end{split}
\end{equation} 
The characteristic equation of the governing equation (\ref{eq:1}) is
\begin{equation}\label{eq:3}
	b_{11}\mu_j^4-2b_{16}\mu_j^3+(2b_{12}+b_{66})\mu_j^2-2b_{26}\mu_j+b_{22}=0.
\end{equation}
By denoting the roots $\mu_i (i=1\sim 4)$($\mu_1=\overline{\mu}_3$ and $\mu_2=\overline{\mu}_4$) and introducing two variables $s_1=\mu_1$, $s_2=\mu_2$, \textit{Airy stress function} for general plane strain problem can be expressed as 
\begin{equation}\label{eq:4}
	U(x,y)=2\Re[U_1(z_1)+U_2(z_2)],
\end{equation}
where $z_i=x+s_iy,(i=1,2)$, $U_1$ and $U_2$ are arbitrary functions to be determined by the boundary conditions. 
Hence, the displacements fields are expressed as
\begin{equation}\label{eq:5}
	\begin{split}
		&u_x=2\Re[p_1\phi(z_1)+p_2\psi(z_2)],\\
		&u_y=2\Re[q_1\phi(z_1)+q_2\psi(z_2)],
	\end{split}
\end{equation}
where $p_i$ and $q_i$ are of the form
\begin{equation}\label{eq:6}
	\begin{split}
		&p_1=b_{11}s_1^2+b_{12}-b_{16}s_1, \ p_2=b_{11}s_2^2+b_{12}-b_{16}s_2,\\
		&q_1=\frac{b_{12}s_1^2+b_{22}-b_{26}s_1}{s_1}, \ q_2=\frac{b_{12}s_2^2+b_{22}-b_{26}s_2}{s_2}.
	\end{split}
\end{equation}
By considering the boundary conditions of a crack in an infinite medium, one could get the stress function $\phi$ and $\psi$. 
Thus the displacement fields at the crack-tip, expressed in polar coordinates, is obtained.
\begin{equation}\label{eq:7}
	\begin{split}
		&u_x=K_I\sqrt{\frac{2r}{\pi}} \Re \left[\frac{1}{s_1-s_2} (s_1p_2\sqrt{\cos\theta+s_2\sin\theta}-s_2p_1\sqrt{\cos\theta+s_1\sin\theta})\right]\\
		&u_y=K_I\sqrt{\frac{2r}{\pi}} \Re \left[\frac{1}{s_1-s_2} (s_1q_2\sqrt{\cos\theta+s_2\sin\theta}-s_2q_1\sqrt{\cos\theta+s_1\sin\theta})\right]
	\end{split}
\end{equation}
The the stress fields are
\begin{equation}\label{eq:8}
	\begin{split}
		\sigma_{xx}=K_I \frac{1}{\sqrt{2\pi r}} \Re \left[\frac{s_1s_2}{s_1-s_2} (\frac{s_2}{\sqrt{\cos\theta+s_2\sin\theta}}-\frac{s_1}{\sqrt{\cos\theta+s_1\sin\theta}})\right],\\
		\sigma_{yy}=K_I \frac{1}{\sqrt{2\pi r}} \Re \left[\frac{1}{s_1-s_2} (\frac{s_1}{\sqrt{\cos\theta+s_2\sin\theta}}-\frac{s_2}{\sqrt{\cos\theta+s_1\sin\theta}})\right],\\
		\sigma_{xy}=K_I \frac{1}{\sqrt{2\pi r}} \Re \left[\frac{s_1s_2}{s_1-s_2} (\frac{1}{\sqrt{\cos\theta+s_1\sin\theta}}-\frac{1}{\sqrt{\cos\theta+s_2\sin\theta}})\right].\\
	\end{split}
\end{equation}

\subsection*{S2.2: Relationship between $G_{\rm c}$ and $K_{\rm Ic}$}

Since the stress and energy approaches in linear elastic fracture mechanics (LEFM) are equivalent, there is a unique relation between energy release rate $G_{\rm c}$ and stress intensity factor $K_{\rm Ic}$\cite{irwin1957analysis}.
 The relation can be derived with the aid of \textit{crack closure method}. 
 First, we consider a Mode-I crack before and after extension distance $da$. 
 Two corresponding coordinates $x-y$ and $x'-y'$ centred at the crack tip are established ($x'=x-da$ and $y=y'$). 
 According anisotropic LEFM crack-tip fields, the normal stress ahead of the crack-tip ($x=x, y=0$) before extension can be obtained by plugging $r=x, \cos\theta=1, \sin\theta=0$ into $\sigma_{yy}$ of equation (\ref{eq:8})
\begin{equation}\label{eq:9}
	\sigma_{yy}=K_{\rm I} \frac{1}{\sqrt{2\pi x}}.
\end{equation}
Assuming crack propagation of a small distance $da$, the displacement along $y$ direction of new crack surfaces ($0\leqslant x \leqslant da$) can be calculated by plugging  $x'=x-da$ into $u_y$ of equation (\ref{eq:7})
\begin{equation}\label{eq:10}
	u_{y}=K_{\rm I}\sqrt{\frac{2(da-x)}{\pi}} \Re \left[\frac{s_1}{s_1-s_2}, (s_1 q_2-s_2 q_1)\right].
\end{equation}
The strain energy associated with this process can be calculated as the work done by $\sigma_{yy}$ before extension, which closes up the crack opening after the crack extension. 
The work done by $\sigma_{yy}$ traversing $u_y$ equals to the energy release rate $G_{\rm I}da$ under mode-I extension, leading to 
\begin{equation}\label{eq:11}
	G_{\rm c}da=2\int_0^{da} \frac{1}{2}\sigma_{yy}u_ydx.
\end{equation}
Substitution of $\sigma_{yy}$ (equation \ref{eq:9}) and $u_{y}$ (equation \ref{eq:10}) into equation (\ref{eq:11}) results in the expression 
\begin{equation}\label{eq:12}
	G_{\rm c}da=\frac{K_{\rm Ic}^2}{2}\Re\left[\frac{s_1}{s_1-s_2} (s_1q_2-s_2q_1)\right] ,
\end{equation}
where $s_1$ and $s_2$ are obtained by solving the governing equation (\ref{eq:3}).
Especially when the material  has orthotropic symmetry with respect to $x-z$ and $y-z$ planes ($a_{16}=a_{26}=0$), equation (\ref{eq:2}) reduces to 
\begin{equation}\label{eq:13}
	b_{11}\mu_j^4+(2b_{12}+b_{66})\mu_j^2+b_{22}=0.
\end{equation}
An analytical solution is thus obtained for $s_1$ and $s_2$
\begin{equation}\label{eq:14}
	\begin{split}
		s_1^2=\mu_1^2=\frac{2b_{12}+b_{66}}{2b_{11}}+\frac{1}{2b_{11}}\sqrt{(2b_{12}+b_{66})^2-4b_{11}b_{22}},\\
		s_2^2=\mu_2^2=\frac{2b_{12}+b_{66}}{2b_{11}}-\frac{1}{2b_{11}}\sqrt{(2b_{12}+b_{66})^2-4b_{11}b_{22}}.
	\end{split}
\end{equation}
$q_1$ and $q_2$ are determined by elastic constants (equation (\ref{eq:6})).
Then the energy release rate reads
\begin{equation}\label{eq:15}
	G_{\rm c}=K_{\rm Ic}^2B,
\end{equation}
where \cite{Liebowitz}
\begin{equation}\label{eq:16}
	B=\sqrt{\frac{b_{11}b_{22}}{2}\left(\frac{2b_{12}+b_{66}}{2b_{11}}+\sqrt{\frac{b_{22}}{b_{11}}}\right)}.
\end{equation}
As discussed in section S2.1, B depends on the elastic constants. Since, according to Griffith theory\cite{Griffith}, $G_{\rm c}=2 \gamma_{\rm s}$, then the Griffith prediction of the critical stress intensity factor $K_{\rm G}$ is
\begin{equation}
	K_{\rm G} = \sqrt{2 \gamma_{\rm s} / B},
\end{equation}
which is equation (1) in the main manuscript.

\subsection*{S2.3: Anisotropic Rice criterion for dislocation emission from crack tip}

Equation (2) of the main manuscript predicts the critical $K_{\rm Ie}$ for dislocation emission from the crack tip \cite{RICE1992239,sun1994dislocation,Andric_2018} 
\begin{equation}
	K_{\rm Ie}=\frac{\sqrt{\gamma_{\rm usf}o(\theta,\phi)}}{F_{12}(\theta)},
\end{equation}
where $\theta$ is the angle between slip plane and crack plane, $\phi$ is the angle between slip direction and a vector lying on the slip plane and perpendicular to the crack-front direction, and $\gamma_{\rm usf}$ is the unstable stacking fault energy of the slip plane.
 $F_{12}(\theta)$ is a geometrical factor 
\begin{equation}
	\frac{K_{\rm I}}{\sqrt{2\pi r}}F_{12}(\theta)=(\sigma_{yy}-\sigma_{xx})\sin\theta\cos\theta+\sigma_{xy}(\cos^2\theta-\sin^2\theta)
\end{equation}
where $\sigma_{ij}$ is the near crack-tip stress field given in equation \ref{eq:8}.
$o(\theta,\phi)$ is an elasticity coefficient 
\begin{equation}
	o(\theta,\phi)=s_i(\phi)\Lambda_{ij}^{-1}(\theta)s_j(\phi)
\end{equation}
where
\begin{equation}
	s(\phi)=(\cos\phi,0,\sin\phi)
\end{equation}
and 
\begin{equation}
	\Lambda_{ij}(\theta)=\Omega_{ik}\Lambda_{kl}\Omega_{jl}.
\end{equation}
$\bm{\Omega}$ is the rotation matrix 
\begin{equation}
	\bm{\Omega}=
\begin{bmatrix}
	\cos\theta & \sin\theta & 0 \\
	-\sin\theta & \cos\theta & 0 \\
	0 & 0 & 1 \\
\end{bmatrix}
\end{equation}
and $\bm{\Lambda}$ is the Stroh tensor \cite{Stroh_1958}
\begin{equation}
	\bm{\Lambda}=\frac{1}{2} \Re(i\mathbf{AB}^{-1}).
\end{equation}
$\mathbf{A}$ and $\mathbf{B}$ can be solved by using the Stroh formalism \cite{Stroh_1958,Ting_1996}
\begin{equation}\label{eq:25}
	\mathbf{N}
	\begin{bmatrix}
		\mathbf{A} \\
		\mathbf{B} \\
	\end{bmatrix}
=\bm{\nu}
\begin{bmatrix}
		\mathbf{A} \\
		\mathbf{B} \\
\end{bmatrix}
.
\end{equation}
Equation \ref{eq:25} is an eigenvalue problem where $\mathbf{A}$ and $\mathbf{B}$ are matrices formed by eigenvectors 
\begin{equation}
	\mathbf{A} = [\bm{a_1},\bm{a_2},\bm{a_3}],\ \\
	\mathbf{B} = [\bm{b_1},\bm{b_2},\bm{b_3}], \\
\end{equation}
and $\bm{\nu}$ is a diagonal matrix filled up by eigenvalues
\begin{equation}
	\bm{\nu}=\rm{diag}[\nu_\alpha].
\end{equation}
$\mathbf{N}$ is the fundamental elasticity matrix
\begin{equation}
	\mathbf{N}=
	\begin{bmatrix}
		\mathbf{N_1} & \mathbf{N_2} \\
		\mathbf{N_3} & \mathbf{N_1}^{\rm T} \\
	\end{bmatrix}
\end{equation}
where
\begin{equation}
	\mathbf{N_1=-T^{-1}R^{T}}, \mathbf{N_2=T^{-1}}, \ \rm{and} \ \mathbf{N_3=RT^{-1}R^T-Q}
\end{equation}
with
\begin{equation}
	\begin{split}
		\mathbf{Q}=       
		\begin{bmatrix}
			C_{11} & C_{16} & C_{15} \\
			C_{16} & C_{66} & C_{56} \\
			C_{15} & C_{56} & C_{55} \\
		\end{bmatrix}
,
		\mathbf{R}= 
		\begin{bmatrix}
			C_{16} & C_{12} & C_{14} \\
			C_{66} & C_{26} & C_{46} \\
			C_{56} & C_{25} & C_{45} \\
		\end{bmatrix}
,\ \rm{and} \
		\mathbf{T}= 
		\begin{bmatrix}  
			C_{66} & C_{26} & C_{46} \\
			C_{26} & C_{22} & C_{24} \\
			C_{46} & C_{24} & C_{44} \\
		\end{bmatrix}
	.
	\end{split} 
\end{equation}
The detailed derivation can be found in chapters 5 and 11 of Ref. \cite{Ting_1996}. 

\section*{S3: Summary of calculated properties for the classical potentials}

Table \ref{table:1} reports the potential-dependent properties that have been used to calculate $K_{\rm G}$ and $K_{\rm Ie}$ for the interatomic potentials considered in Fig. 1 of the main manuscript.

\begin{table}[H]
	\centering
	\caption{Summary of calculated properties for the classical potentials used in Fig. 1. ($a_0$: lattice constant. 
		$C_{\rm ij}$ : elastic constants, 
		$\gamma_{\rm s}$: surface energies, 
		$\gamma_{\rm usf}$ : unstable stacking fault energies)}
	\begin{tabular}{c|c|c|c|c|c|c|c} 
		\hline
		\hline
		Properties & Zhou\cite{zhou_2004_PhysRevB.69.144113} &  Olsson\cite{olsson_2009} & Li\cite{meam_liyange_PhysRevB.89.094102} & Asadi\cite{Asadi_2015} & Etesami\cite{Asadi_2018} & Byggmästar\cite{Byggmastar_2020} & Starikov\cite{adp_PhysRevMaterials.5.063607} \\  [0.5ex]
		\hline
		\hline
		Potential & EAM &EAM & MEAM & MEAM & MEAM & BOP & ADP \\
		\hline
		$a_0$ & 2.866 & 2.87 & 2.851 & 2.851 & 2.851 & 2.860 & 2.83\\
		\hline
		$C_{11}$ & 230 & 239 .55& 228.25 &231.27 & 254.05 & 222.90 & 255 \\
		\hline
		$C_{12}$ & 136 & 135.75 & 135.39 & 134.5 & 123.12 & 145.40 & 116\\
		\hline
		$C_{44}$ & 117 & 120.75 & 117.97 & 116.19 & 126.84 & 117.9 & 113\\
		\hline
		$\gamma_{\rm s}\{100\}$ & 1.689 & 1.621 & 1.940 & 2.360 & 2.452 & 1.677 & 2.636\\
		\hline
		$\gamma_{\rm s}\{110\}$ &1.429 & 1.482 & 2.143 & 2.3647 & 2.3195 & 1.349 & 2.421 \\
		\hline
		$\gamma_{\rm usf}\{110\}$ & 0.7161 &0.8115 &0.6424 &0.8056 &0.9092 &0.8368 &0.8238 \\
		\hline
		$\gamma_{\rm usf}\{112\}$ &0.8315 &0.9457& 0.6434 &1.653 &1.757 &0.9634 &1.0323 \\
		\hline
		\hline
	\end{tabular}
	\label{table:1}
\end{table}

\section*{S4: SOBOL sampling approach}

First, a set of 9 dimensional data (the primitive cell parameters) is generated and the strains of primitive cells are calculated.
The strain components that fall into the range predicted by LEFM are then selected, and data within a norm distance of 0.2 (unit-less) in comparison with the original database are excluded. 
The data exclusion is performed in order to reduce repeating information of the entire database.
To enrich the potential energy surface (PES) description with highly strained states, we created 1744 highly-distorted primitive bcc cell, as follows: \\
\indent (i) SOBOL sampling 4,000 data points in 9 dimensional space; \\
\indent (ii) Select the data whose strain components are confined to the range of $[-0.22,0.3]$, i.e. 2427 points; \\
\indent (iii) Calculate the norm between the new points and points from the original database; \\
\indent (iv) Set a criterion (0.2 in this study) and exclude the data of new points if the norm is smaller than this criterion; \\
\indent (v) Perform DFT calculation of selected primitive cells; \\
\indent (vi) Exclude the non-magnetic data. 

\section*{S5: Principal component analysis of DB1 and DB9}

\begin{figure}[H]
	\centering
	\includegraphics[trim=0 10 0 40, clip, scale=0.4]{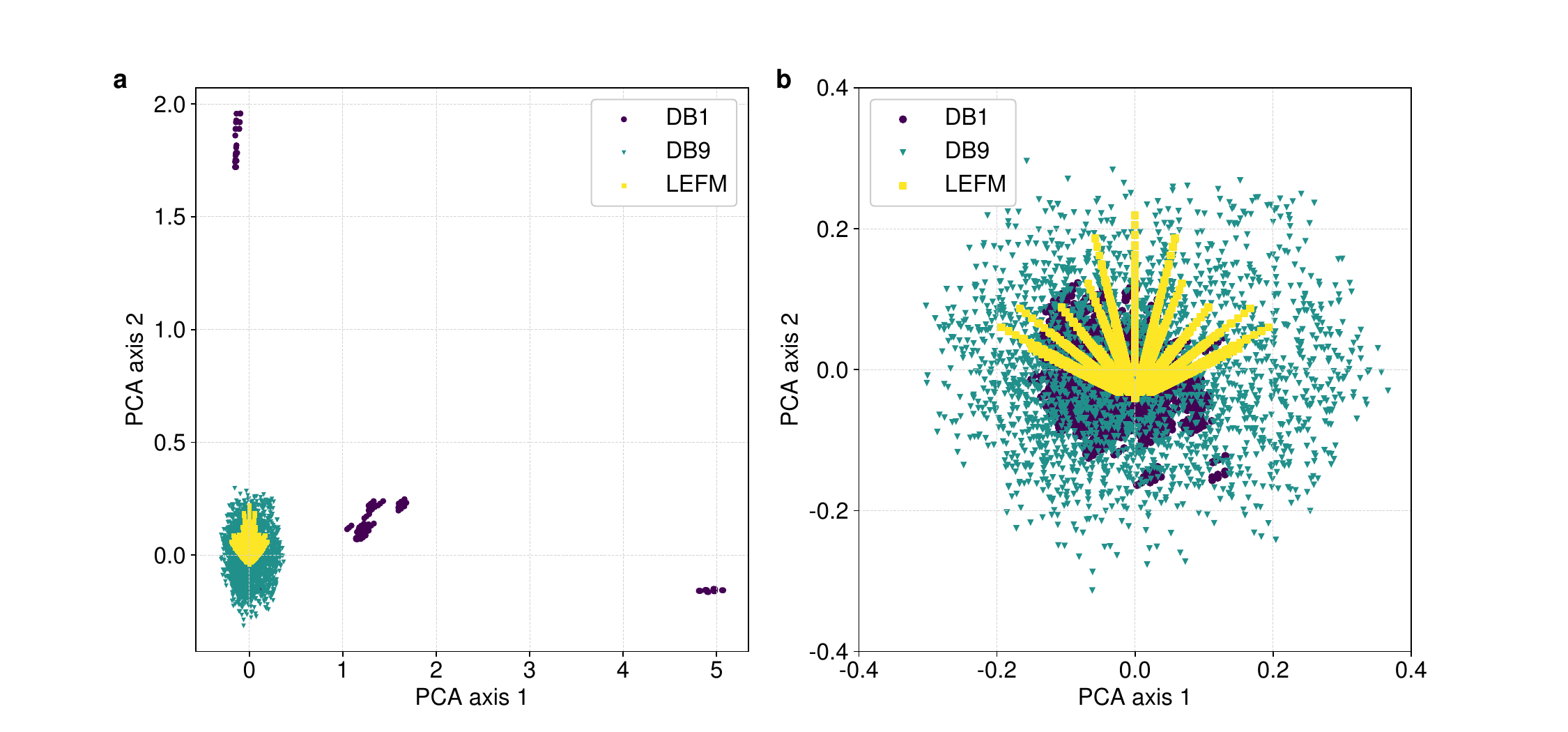}
	\caption{\textbf{(a)} Principal component analysis of strain components from DB1 (original database), DB9 and LEFM.
		\textbf{(b)} Zoom-in of the data that is centred in the origin of \textbf{(a)}.}
	\label{fig:pca}
\end{figure}

Fig. \ref{fig:pca}\textbf{a} plots the first two components of the principal component analysis results of the strain components from DB1, DB9 and LEFM. 
The plot helps visualise the distribution of the primitive cell configuration in the 6-dimensional strain space.
The data that is far from the original represents the cells that have large distortion along at least one or two crystallographic directions. 
LEFM data is obtained under mode-I loading from $K_{\rm I}=0.7\ \rm MPa\sqrt{m}$ to $K_{\rm I}=1.5\ \rm MPa\sqrt{m}$. 
Fig. \ref{fig:pca}\textbf{b} zooms into the main data of Fig. \ref{fig:pca}\textbf{a}.
DB9 expands the boundary of DB1, and LEFM data occupies specific directions in the PCA space. 
The plot shows that SOBOL sampling encompasses the highly symmetric LEFM configurations, and hence it incorporates crack-relevant data.

\section*{S6: Magnetic moment of crack-tip atoms}

\begin{figure}[H]
	\centering
	\includegraphics[trim=0 0 0 0, clip, scale=0.4]{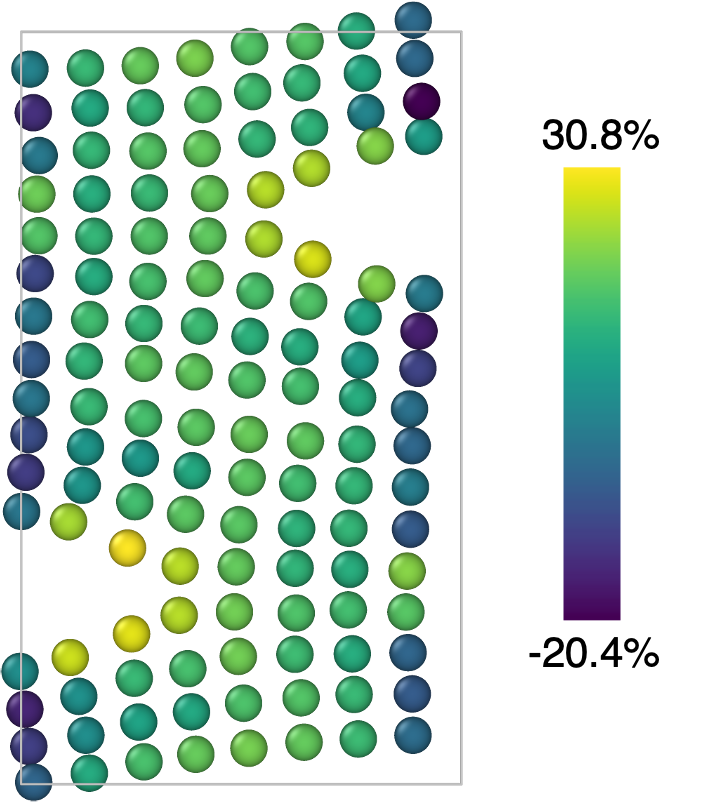}
	\caption{Magnetisation degree of a DFT crack-tip cell. Atoms are coloured by their magnetic moment change with respect to the perfect bulk bcc ferromagnetic iron.}
	\label{fig:dft_mag}
\end{figure}

We performed DFT calculations of approximately periodic crack-tip cells. The magnetic moment change of each atom with respect to bulk bcc ferromagnetic iron is plotted in Fig. \ref{fig:dft_mag}. We can see that the atoms at crack-tip remain in the ferromagnetic state, with a maximum local magnetisation change of $\sim$30\%.

\vspace*{10em}
\section*{S7: Information about the preliminary database}


Table \ref{tab:db_info} lists the information related to the preliminary, crack-relevant database (``\textit{Iter-0}" GAP).

\begin{table}[H]
	\centering
	\caption{Information of DB9, DB10 and DB11.}
	\begin{tabular}{c|c|c|c|c|c|c}
		\hline\hline
		\multirow{2}{*}{} & Target & Total number & $N$ of atoms & Simulation & $k$ spacing & Plane\\
		& property & of LAEs & in DFT cell & box & $1/2\pi\rm\AA^{-1}$ &  \\
		\hline\hline
		\multirow{2}{*}{DB9} & highly & \multirow{2}{*}{1744} & \multirow{2}{*}{1} & distorted  & \multirow{2}{*}{0.015} & \multirow{2}{*}{-}\\ 
		& deformed bulk &  & & bcc primitive &  & \\
		\hline
		\multirow{2}{*}{DB10} & highly      & \multirow{2}{*}{396}  & \multirow{2}{*}{2} & distorted  & \multirow{2}{*}{0.015} & \multirow{2}{*}{-}\\ 
		& deformed bulk &  & & hcp primitive &  & \\
		\hline
		\multirow{4}{*}{DB11} &  & 528 & 24&  $1\times 1\times 24$ & 0.03& $(100)$ \\
		& surface    & 440           & 20&  $1\times 1\times  20$ & 0.03& (110)\\
		& separation & 1056          & 48&  $1\times 2\times  24$ & 0.03& (111)\\
		&            & 660           & 30&  $1\times 2\times  15$ & 0.03& (112)\\
		\hline\hline
	\end{tabular}
	\label{tab:db_info}
\end{table}

\section*{S8: Fracture prediction by GAP trained on preliminary database (``\textit{Iter-0}")}

\begin{figure}[H]
	\centering
	\includegraphics[trim=0 10 0 40, clip, scale=0.4]{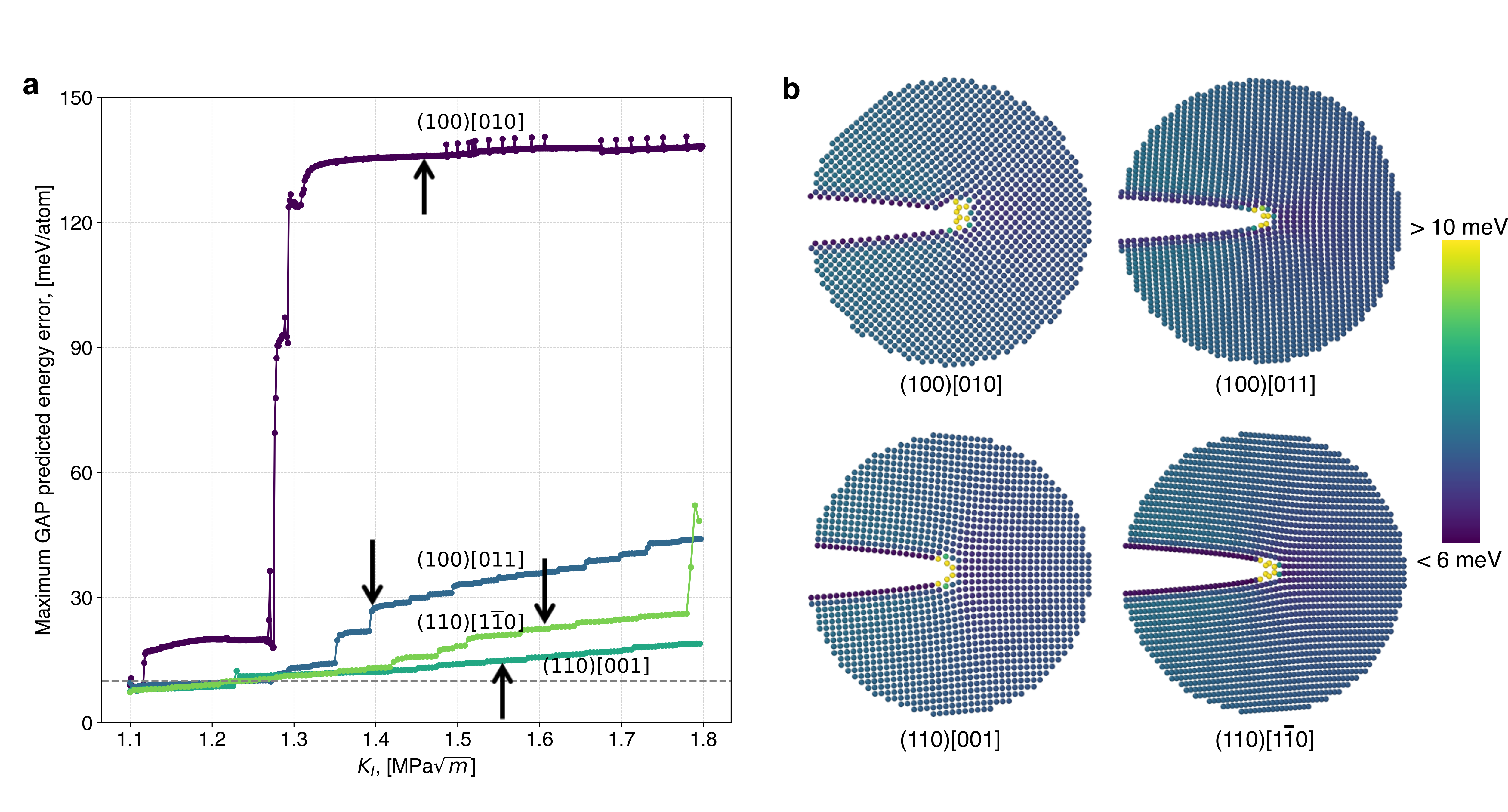}
	\caption{Fracture predictions of GAP trained on preliminary database, i.e. DB9, DB10 and DB11. \textbf{(a)} Maximum per-atom GAP predicted energy error during fracture simulations for four crack systems. The dashed line indicates the 10 meV/atom, and the arrows indicate the configuration shown in \textbf{(b)}. \textbf{(b)} Simulation snapshots coloured by GAP predicted per-atom energy error. }
	\label{fig:gap_error_manual}
\end{figure}

\section*{S9: Convergence of original SOAP}

\begin{figure}[H]
	\centering
	\includegraphics[scale=0.45]{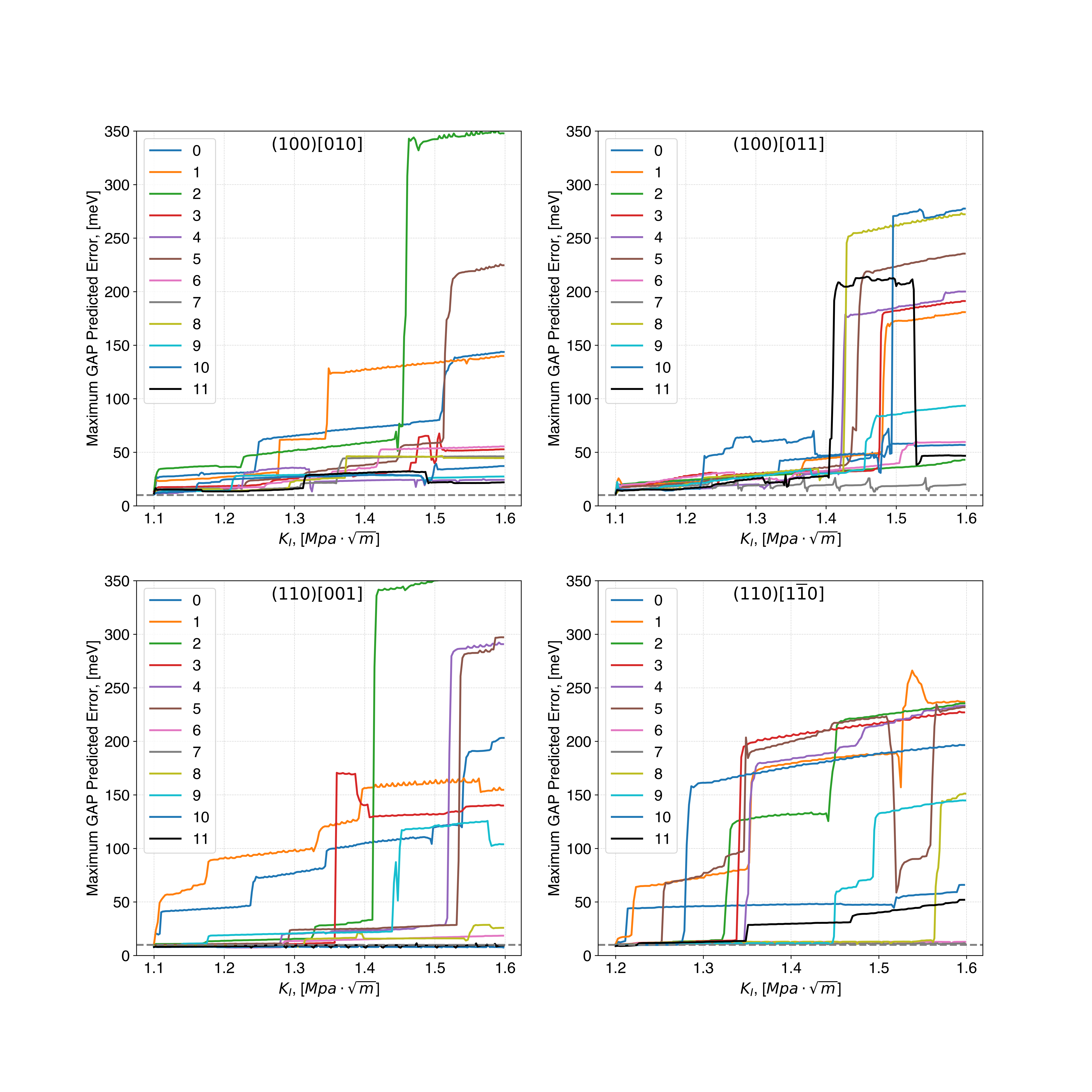}
	\caption{Convergence analysis of GAP trained on original SOAP. Maximum GAP predicted per-atom energy error as a function of applied $K_{\rm I}$ during the fracture simulation for four crack systems.}
	\label{fig:gap_error_normal_soap}
\end{figure}

\begin{figure}[H]
	\centering
	\includegraphics[trim=0 20 0 80, clip, scale=0.45]{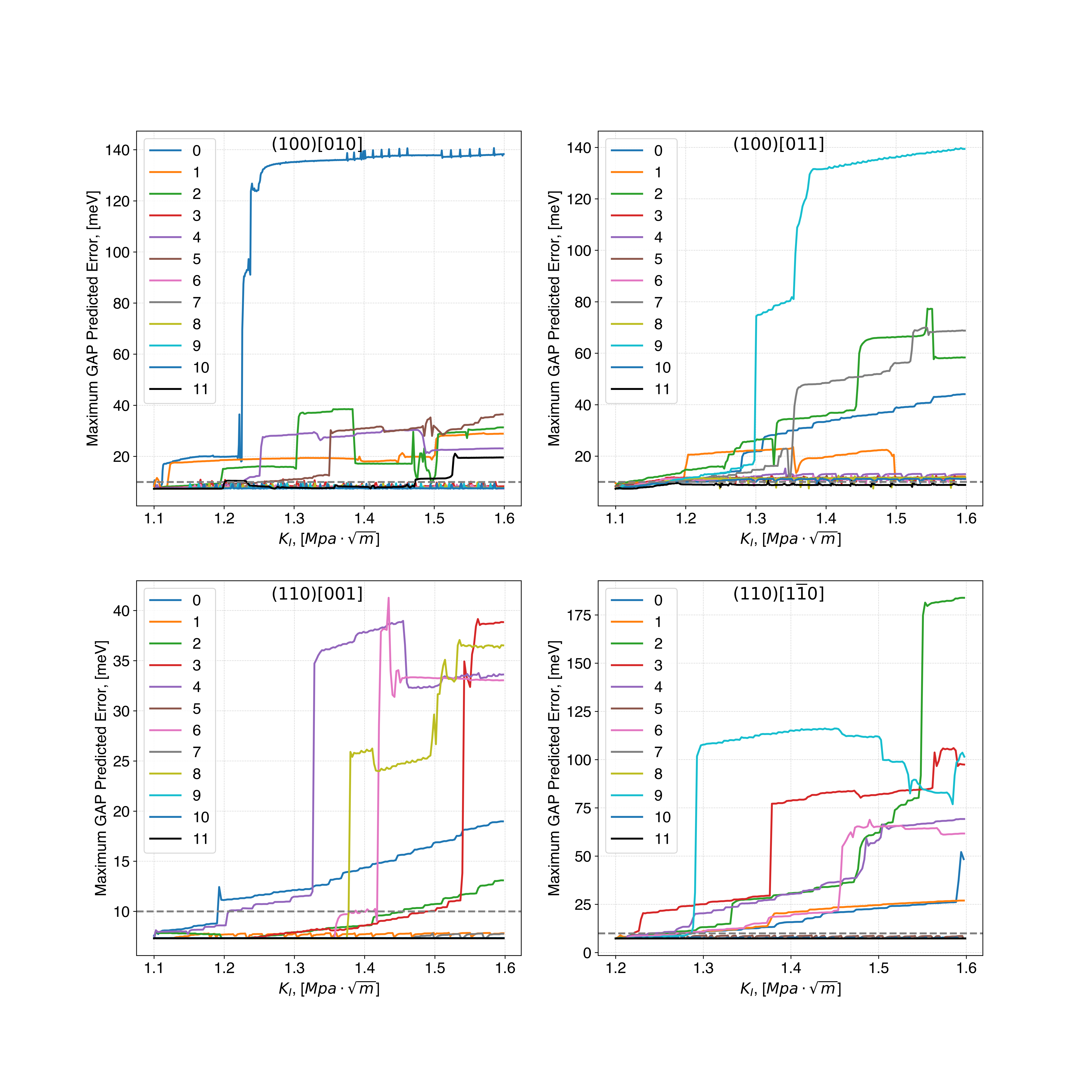}
	\caption{Convergence analysis of GAP trained on turbo SOAP by using the database developed by original SOAP.
	Maximum GAP predicted per-atom energy error as a function of applied $K_{\rm I}$ during the fracture simulation for four crack systems.}
	\label{fig:gap_error_turbo_soap}
\end{figure}

We also train GAP with the original SOAP descriptor on crack system $(110)[001]$. 
The GAP predicted variance of $(110)[001]$ converges to 10 meV/atom after 11 iterations, as shown in Fig. \ref{fig:gap_error_normal_soap}.
However, other three crack systems are not converged yet. 
Since Turbo-SOAP descriptor has a set of better performing basis functions compared with SOAP, we train a series of GAPs with Turbo-SOAP based on the dataset developed by using original SOAP, referred to as Fe-S-GAP22. 
We found a better convergence in terms of GAP predicted per-atom energy error, as shown in Fig. \ref{fig:gap_error_turbo_soap}.
Except that the error of $(100)$ crack systems shows slight fluctuations, the predicted errors are converged to less than 10 meV/atom for all crack systems .
The predicted fracture mechanisms are the same as obtained with Turbo SOAP, suggesting the fracture mechanism is independent of database as long as the GAP predicted error is converged.

\section*{S10: Benchmark of GAP22}

\begin{figure}[H]
\centering
\includegraphics[trim=0 20 0 40, clip,scale=0.4]{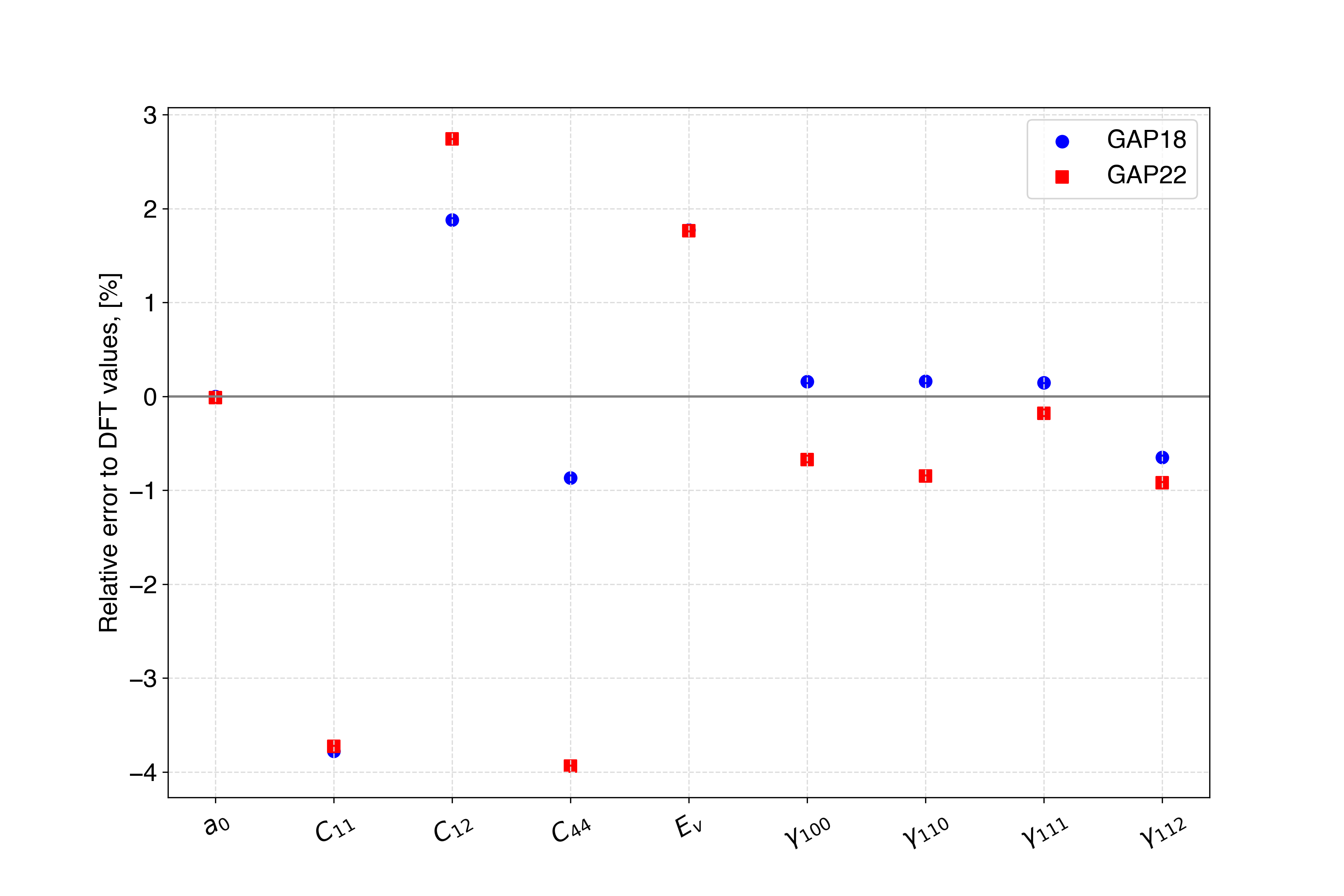}
\caption{Relative errors for various material properties with respect to DFT of Fe-GAP18 and Fe-GAP22. In particular, lattice parameter ($a_0$), elastic constant ($C_{11},C_{12},C_{14}$), vacancy formation energy ($E_v$) and surface energies ($\gamma_{100},\gamma_{110},\gamma_{111},\gamma_{112}$) are considered.}
\label{fig:gap_test}
\end{figure}

Fig. \ref{fig:gap_test} shows that the prediction of elastic constants of Fe-GAP18 and Fe-GAP22 are in good agreement with each other. $\rm C_{44}$ predicted by Fe-GAP22 is 4\% off with respect to DFT. The predicted surface energies of \{100\} and \{110\} planes have a maximum difference of 1\%.  
\vspace*{-2em}
\section*{S11: Normal stress distribution of the crack tip solved by anisotropic LEFM}

\begin{figure}[H]
	\centering
	\includegraphics[trim=0 20 0 40, clip, scale=0.4]{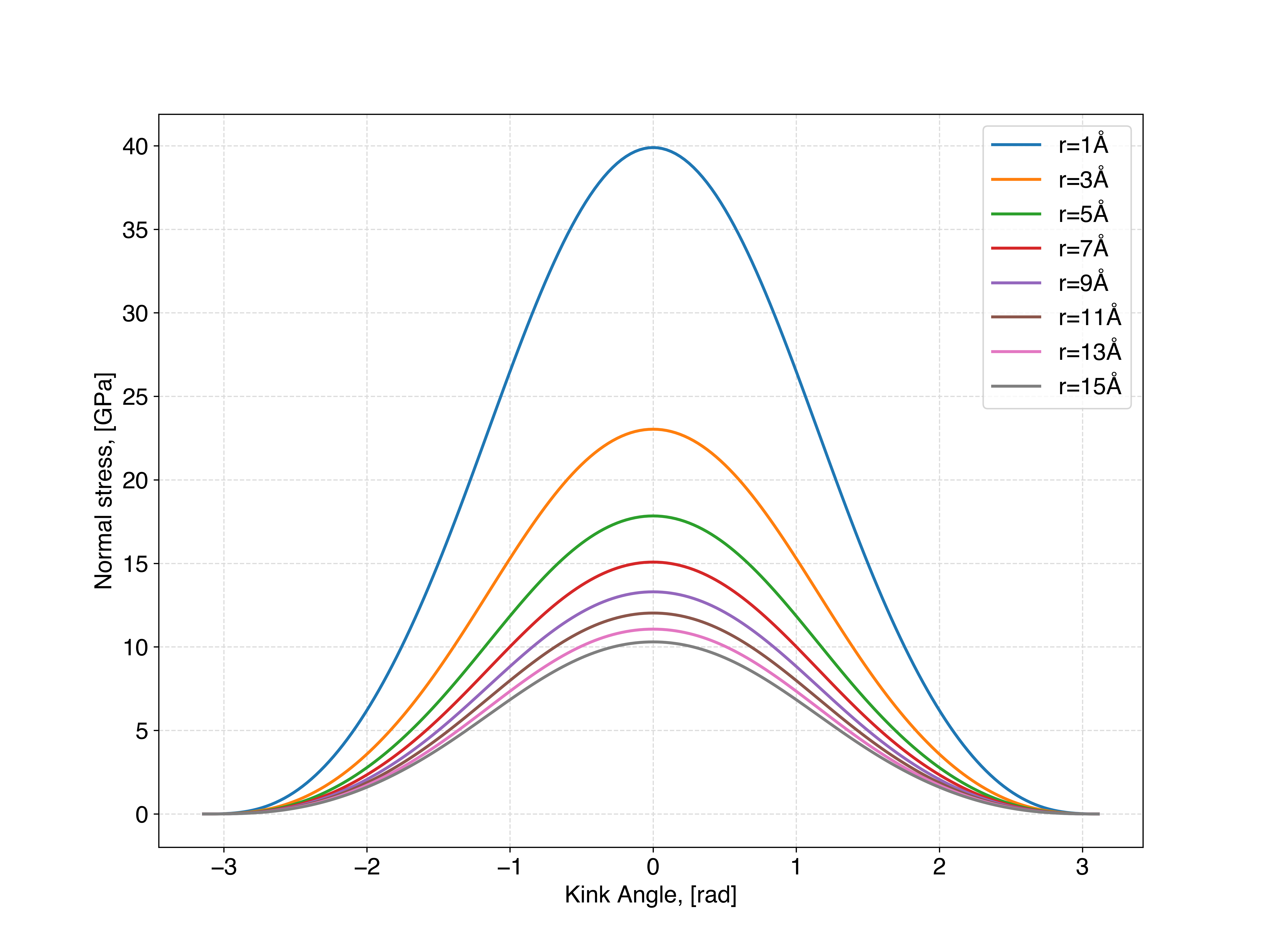}
	\caption{Normal stress ($\sigma_{\theta\theta}$) distribution around the crack tip computed using anisotropic LEFM. $r$ is the radius.}
	\label{fig:normal_stress}
\end{figure}

Fig. \ref{fig:normal_stress} plots the normal stress distribution as a function of the angle with respect to the crack plane for radius from 1 Å to 15 Å. 
The normal stress $\sigma_{\theta\theta}$ is calculated from equation (\ref{eq:16}). 
Anisotropic LEFM (see \textit{Section S.2}) predicts the maximum stress normal to the crack plane.

\section*{S12: GAP22 predicted error at T=300K during fracture simulation}

\begin{figure}[H]
	\centering
	\includegraphics[scale=0.55]{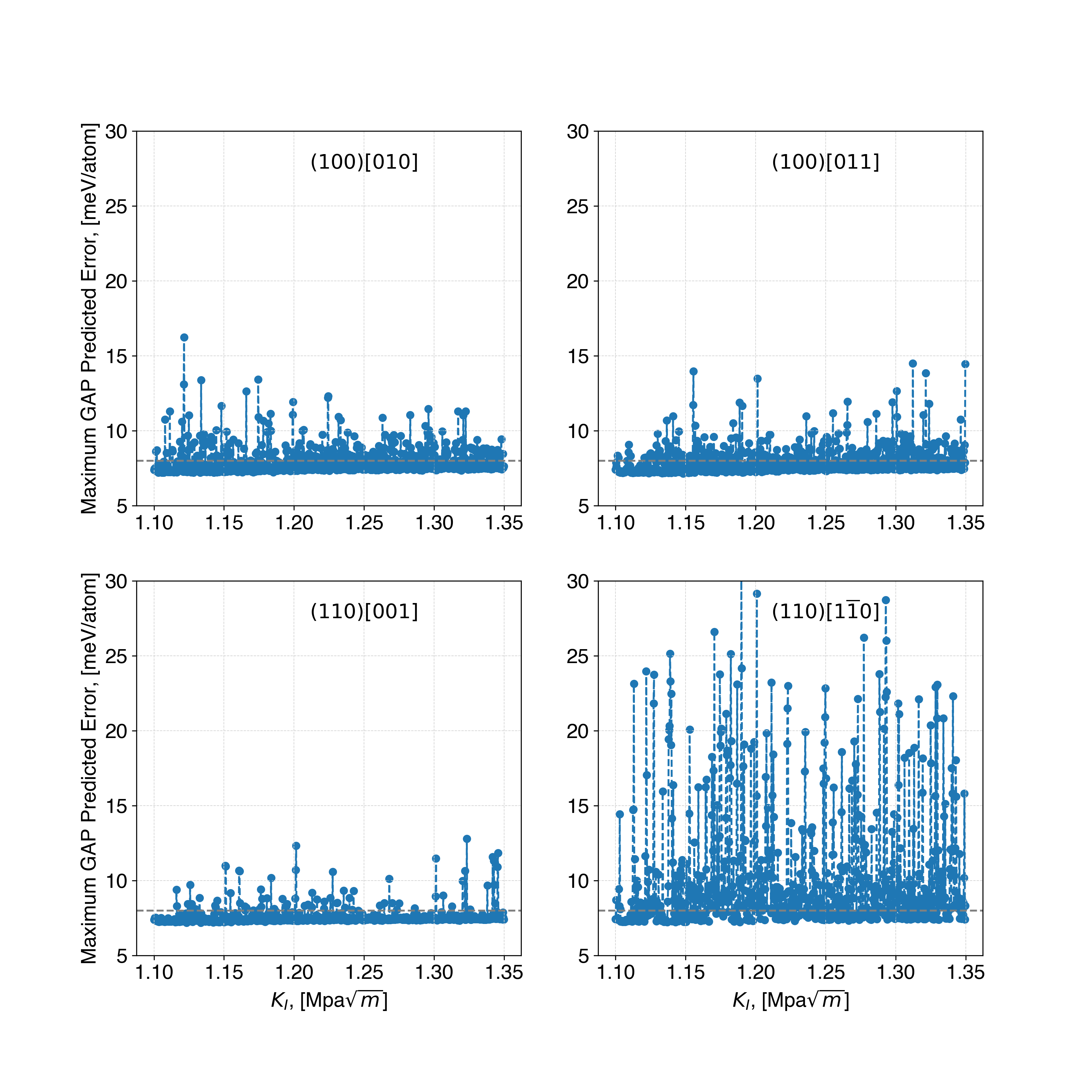}
	\caption{Maximum GAP predicted per-atom energy error as functions of $K_{\rm I}$ for four crack systems at T=300K: (a)$(100)[011]$, (b)$(110)[001]$, (c)$(100)[010]$ and (d)$(110)[1\bar{1}0]$. The dashed line indicates 8 meV/atom.}
	\label{fig:gap_error_300K}
\end{figure}

We performed fracture simulation of four crack systems at T=300K under a loading rate of $10^9 \rm MPa\sqrt{m}s^{-1}$. 
Fig. \ref{fig:gap_error_300K} plots the maximum GAP prediction energy error per-atom as a function of $K_{\rm I}$ for four crack systems. 
For all crack system expect for $(110)[1\overline{1}0]$, the maximum error increases to 15 meV/atom due to the thermal fluctuation. 
For $(110)[1\overline{1}0]$, the maximum error is around 30 meV/atom.

\section*{S13: $K_{\rm IC}$ predicted by Fe-S-GAP22}

\begin{figure}[H]
	\centering
	\includegraphics[scale=0.4]{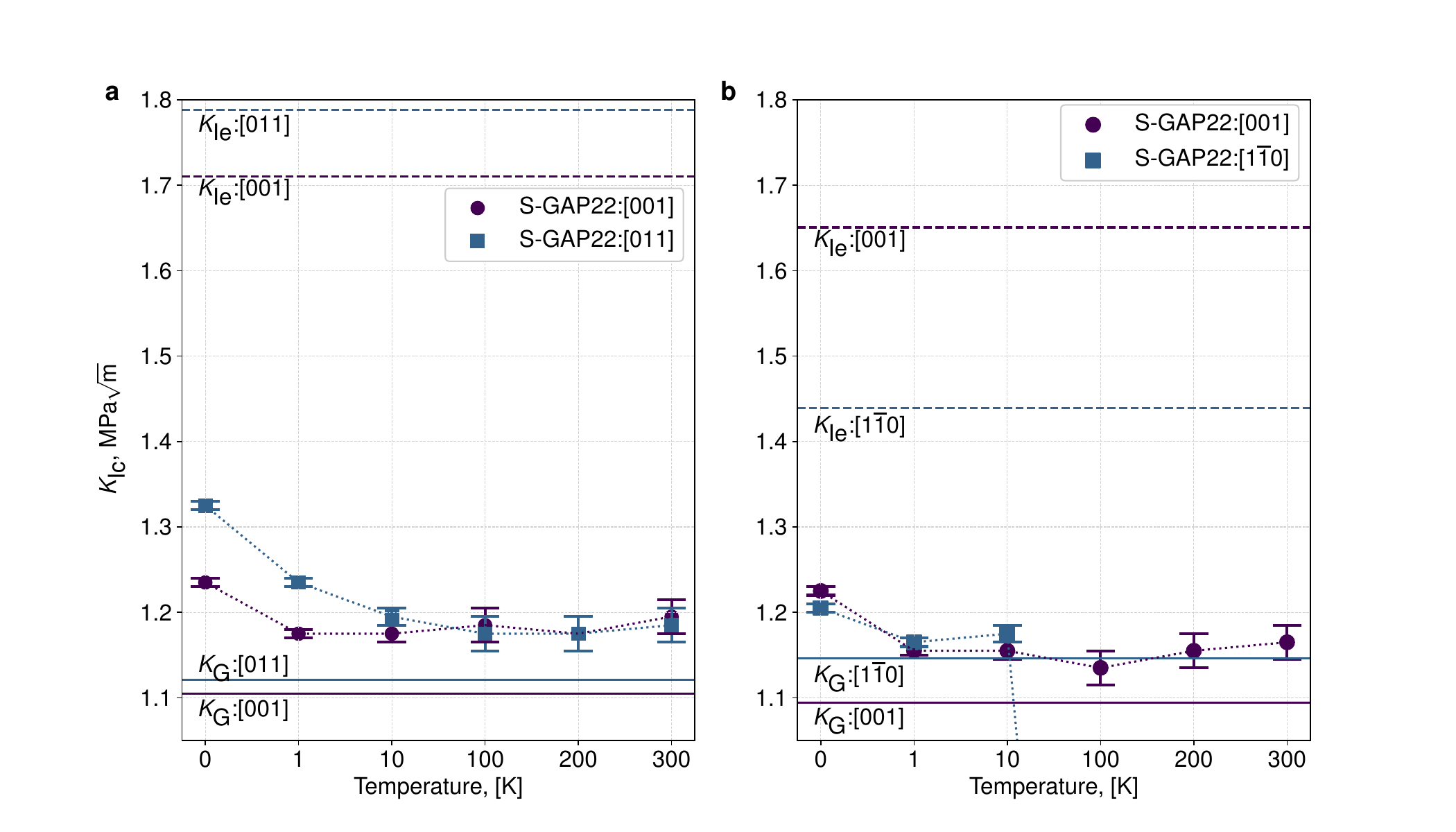}
	\caption{Critical $K_{\rm I}$ predicted by Fe-S-GAP22 at different temperature (T=0K-300K). 
		\textbf{(a)}(100) and \textbf{(b)}(110) crack plane. 
		The solid and dashed line indicate critical $K_{\rm I}$ predicted by Griffith ($K_{\rm G}$) and Rice ($K_{\rm Ie}$) theories, respectively.}
	\label{fig:soap_kic}
\end{figure}

Fig. \ref{fig:soap_kic} shows the critical $K_{\rm I}$ predicted by Fe-S-GAP22.
The predicted $K_{\rm Ic}$ decreases from T=0K to 1K, which is caused by the breaking of symmetry and overcoming local minute barriers.
The predicted $K_{\rm Ic}$ converges with respected to temperature, showing the similar trend as the prediction of Fe-GAP22.

\section*{S14: Schematic plot of MS/MD simulation setup}

\begin{figure}[H]
	\centering
	\includegraphics[scale=0.4]{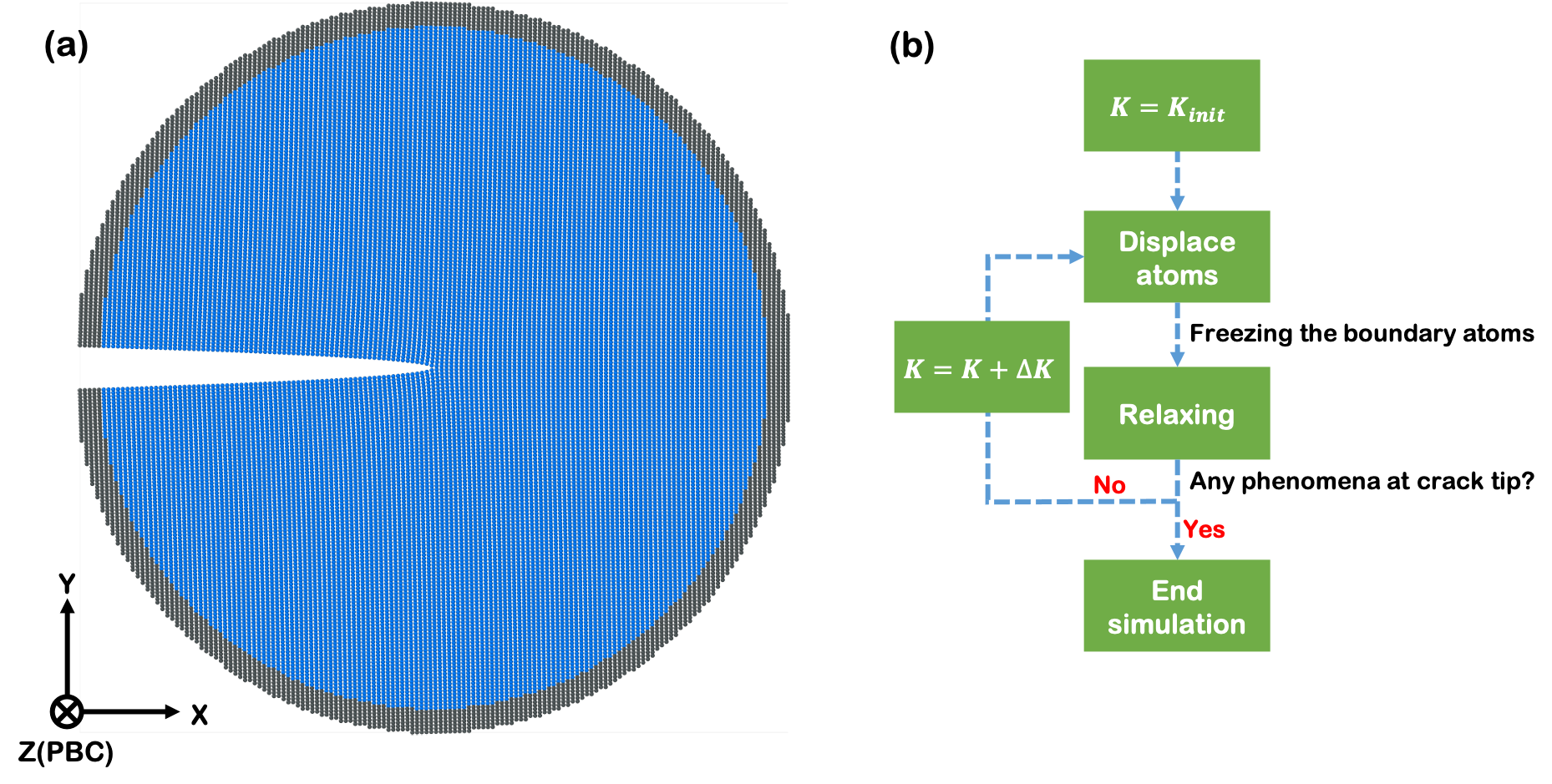}
	\caption{\textbf{(a)}Schematic plot of atomistic fracture simulation setup. $x$, $y$ and $z$ are aligned with the crack propagation direction, crack plane normal and crack front, respectively.
	\textbf{(b)} $K$-test iteration algorithm. }
	\label{fig:md_setup}
\end{figure}

Fig. \ref{fig:md_setup}\textbf{a} plots the $K$-test specimen used for atomistic fracture simulation. During the loading, displacements are assigned to the grey boundary atoms. The mobile blue atoms are subsequently either relaxed (in MS) or equilibrated (in MD). As shown in Fig. \ref{fig:md_setup}\textbf{b}, the $K$ displacement fields are applied incrementally until the crack-tip event takes place.  

\bibliographystyle{unsrt}
\bibliography{supp}